\documentclass{PoS}
\newcommand{\beq}{\begin{equation}}
\newcommand{\eeq}{\end{equation}}
\newcommand{\beqa}{\begin{eqnarray}}
\newcommand{\eeqa}{\end{eqnarray}}
\usepackage{bm}

\title{High-energy Emission Properties of Pulsars}

\ShortTitle{High-energy Emission Properties of Pulsars}

\author{\speaker{Christo Venter},$^a$ Alice K.\ Harding$^{b}$
and Isabelle Grenier$^{c}$\\
\llap{$^a$}Centre for Space Research, North-West University, Potchefstroom Campus, Private Bag X6001, Potchefstroom 2520, South Africa\\
\llap{$^b$}Astrophysics Science Division, NASA Goddard Space Flight Center, Greenbelt, MD 20771, USA\\
\llap{$^c$}Laboratoire AIM, CEA-IRFU/CNRS/Universit\'e Paris Diderot, Service d'Astrophysique, CEA Saclay, 91191, Gif-sur-Yvette, France\\
E-mail: \email{Christo.Venter@nwu.ac.za}, \email{ahardingx@yahoo.com}, \email{isabelle.grenier@cea.fr}}

\abstract{The sheer number of new $\gamma$-ray pulsar discoveries by the \textit{Fermi} Large Area Telescope since 2008, combined with the quality of new multi-frequency data, has caused a revolution in the field of high-energy rotation-powered pulsars. These rapidly rotating neutron stars exhibit rich spectral and temporal phenomenology, indicating that there are still many unsolved mysteries regarding the magnetospheric conditions in these stars -- even after 50 years of research! Indeed, 2017 marks the golden anniversary of the discovery of the first radio pulsar, and theorists and observers alike are looking forward to another half-century of discovery, with many new experiments coming online in the next decades. In this review paper, we will briefly summarise recent HE pulsar observations, mention some theoretical models that provide a basic framework within which to make sense of the varied measurements, and finally review some of the latest theoretical developments in pulsar emission modelling.}

\FullConference{XII Multifrequency Behaviour of High Energy Cosmic Sources Workshop\\
		12-17 June, 2017\\
		Palermo, Italy}

\begin{document}

\section{Introduction -- High-Energy Pulsars}
Pulsars are seen across the electromagnetic spectrum. Their light curves vary with energy and time, but radio light curves\footnote{The $\gamma$-ray fluxes are so low that single pulses are not available, only stacked / phase-averaged ones are feasible.} averaged over pulsation period are usually quite stable. Their spectra span a very wide range in energy, making these rotating neutron stars true multi-frequency objects. The era before the launch of the \textit{Fermi} Large Area Telescope (LAT) was characterised by a mere 7 $\gamma$-ray pulsars that were detected at high confidence \cite{Thompson2004}. However, after nearly ten years in orbit, scanning the full sky in the high-energy (HE) band from $\sim20$~MeV to over 300~GeV~\cite{Atwood2009}, the \textit{Fermi} LAT has now detected over 200\footnote{At the time of writing, there are 209 public \textit{Fermi} pulsars, including 102 millisecond pulsars (MSPs), 76 binaries, 24 black widows, 7 redbacks, 63 young radio-quiet, and 51 young radio-loud pulsars.} $\gamma$-ray pulsars\footnote{https://confluence.slac.stanford.edu/display/GLAMCOG/Public+List+of+LAT-Detected+Gamma-Ray+Pulsars}. The sheer increase in pulsar number enables us to perform population studies, as well as scrutinise temporal and spectral properties of individual objects at an ever increasing level of detail. In this review, we summarise the status of HE observations (Section~\ref{sec:Obs}), describe the basic theoretical framework of HE pulsar physics (Section~\ref{sec:Frame} and~\ref{sec:OldTheory}), and then focus on some new theoretical developments in the field (Section~\ref{sec:Theory}) before we provide a future outlook (Section~\ref{sec:Conc}). For more comprehensive recent reviews on HE pulsars, see Venter~\cite{Venter16_HEASA}, Cerutti and Beloborodov \cite{Cerutti17a}, and references therein. 

\section{HE Observational Breakthroughs -- What Do We See?}\label{sec:Obs}
\begin{figure}[t]
  \begin{center}
  \includegraphics[width=15cm]{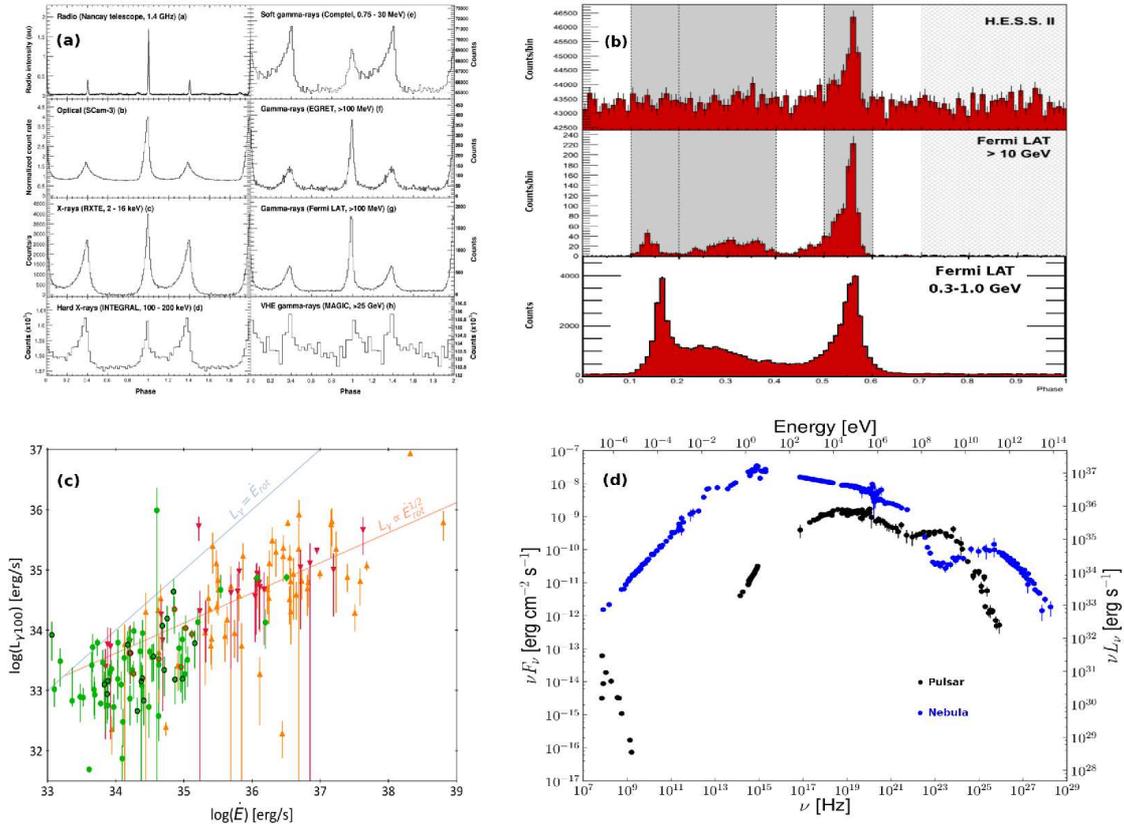}
  \caption{\textit{Panel~(a):} Multi-frequency and subband evolution of the light curves of the Crab pulsar~\cite{Fermi_Crab}. \textit{Panel~(b):} Disappearance of the Vela pulsar's first peak with energy as seen by \textit{Fermi}~\cite{Vela2} and H.E.S.S.~\cite{DeNaurois15}. \textit{Panel~(c):} Updated plot of $L_\gamma$ vs.\ $\dot{E}_{\rm rot}$ for old and young pulsars, exibiting two distinct trends: $L_\gamma\propto\dot{E}_{\rm rot}^{1/2}$ for young pulsars (orange and red dots), while $L_\gamma\propto\dot{E}_{\rm rot}$ for MSPs (green dots)~\cite{Grenier15}. \textit{Panel~(d):} Broad spectral energy distribution of the Crab pulsar (black) and nebula (blue), with the GeV pulsar component showing the typical flat spectrum and exponential cutoff~\cite{Buehler14}.}
  \label{fig:medley}
  \end{center}
\end{figure}

It was clear since the days of the \textit{Compton Gamma-Ray Observatory (CGRO)} that~\cite{Thompson2004}:
\begin{enumerate}
  \item Multi-frequency pulsar light curves are energy-dependent.
  \item $\gamma$-ray pulsar light curves typically exhibit a double-peaked morphology.
  \item The leading pulse typically fades in brightness relative to the trailing pulse as energy is increased.
  \item HE pulsars seem to be relatively young (compared with the full radio population) and to possess large spin-down power $\dot{E}_{\rm rot}=I\Omega\dot{\Omega} = -4\pi^2I\dot{P}/P^3$ (or open-field-line\footnote{Open and closed-field-line regions are separated by the last open $B$-field lines (separatrix) that touch the light cylinder at radius $R_{\rm LC} = c/\Omega$, where the corotation speed is equal to the speed of light.} voltage $V\propto\dot{E}_{\rm rot}^{1/2}$ or particle current $I_{\rm PC}\propto\dot{E}_{\rm rot}^{1/2}$ from the polar cap (PC)), with $I$ the moment of inertia, $\Omega$ the angular frequency, $\dot{\Omega}$ the time derivative thereof, $P$ the period, and $\dot{P}$ its time derivative.
  \item The inferred $\gamma$-ray luminosities of young HE pulsars follow the trend $L_\gamma\propto\dot{E}_{\rm rot}^{1/2}$.
  \item The radiative power in the GeV $\gamma$-ray band (or sometimes soft $\gamma$-ray band, 100 keV -- 1 MeV) dominates the multi-frequency spectrum.
  \item The HE spectra are typically quite hard and typically exhibit spectral cutoffs $E_{\rm cut}$ around a few GeV. Furthermore, in the \textit{CGRO} era, the GeV spectrum of the Vela pulsar was consistent with expectations of both the near-surface PC and high-altitude outer gap (OG) models (See Section~\ref{sec:OldTheory}).
%  \item The spectral cutoff $E_{\rm cut}$ may be related to the neutron star's surface $B$-field $B_0$;
  \item No pulsed TeV emission from pulsars could be detected~\cite{Schmidt05}.
  \item The \textit{Fermi} (formerly GLAST) Mission was expected to find tens to hundreds~\cite{Gonthier04,Gonthier07,Watters11} of HE pulsars, both radio-loud and radio-quiet, aided by its potential for blind period searches using $\gamma$-ray data only. Only very few MSPs were expected to be seen in $\gamma$ rays.
\end{enumerate}

The \textit{Fermi} LAT has confirmed all these basic observational trends (as for number 7, \textit{Fermi} has shown that the emission must originate in the outer magnetosphere, from OGs, SGs or the current sheet), and also confirmed the detection of the 7 high-confidence \textit{CGRO} pulsars (Crab, Vela, B1509$-$58, B1706$-$44, B1951+32, Geminga, and B1055$-$52) as well as the 3 pulsars detected at lower significance (B1046$-$58, B0656+14, and J0218+4232), in addition to more than 200 new HE pulsar detections. Indeed, and additionally, \textit{Fermi} showed that (cf.~\cite{2PC}, Figure~\ref{fig:medley}):
\begin{enumerate}
  \item Pulsar light curves are energy-dependent, with the pulse shapes not only changing for different energy domains, but also for different subbands within the $\gamma$-ray band  (e.g.,~\cite{Vela2}).
  \item HE pulsar light curves often exhibit double-peaked morphology (which generally differ in morphological details, e.g., rapidity of rise and fall of inner / outer peak structures), although there are more complex profiles (e.g., triple peaks and broad or sharp single peaks) as well; furthermore, the radio pulse may be either leading the $\gamma$-ray pulse in phase, be aligned with the $\gamma$-ray peaks, or trailing the $\gamma$-ray pulse~\cite{Johnson14}. There is an inverse trend between the $\gamma$-ray peak separation $\Delta$ and the radio-to-$\gamma$ phase lag $\delta$~\cite{2PC}, as first noted~\cite{RY95} in the context of outer-magnetosphere models with caustic pulses, but which is also predicted in later models involving the current sheet or the beginning of the striped wind~\cite{Kalapotharakos14,Petri11} (Section~\ref{sec:OldTheory}).
  \item For most pulsars, the first peak fades in brightness relative to the second peak with increasing energy, with the Vela and Crab pulsars providing prime examples~\cite{Vela2,MAGIC_Crab12}. However, some pulsar light curves (about one third) show the reverse behaviour~\cite{Brambilla15,Renault15}. In addition, the main peak positions seem to remain more or less the same as the energy increases (but not the third peak of Vela, which migrates in phase with an increase in energy~\cite{Vela2}), while the pulse widths become narrower~\cite{MAGIC_Crab12,DeNaurois15}.
  \item HE pulsars represent the most energetic subpopulation of pulsars in terms of $\dot{E}_{\rm rot}$, with a possible empirical ``death line''\footnote{This means a line in $P\dot{P}$-space below which no pulsar has been detected to date.} occurring at $\dot{E}_{\rm rot}\sim10^{33}$~erg\,s$^{-1}$~\cite{Guillemot16}, set by the much older population of MSPs that has now been detected. At the highest spin-down powers (e.g., PSR B1509$-$58), the spectrum may cut off in the $1-100$~MeV range~\cite{Kuiper15}.
  \item The inferred $\gamma$-ray luminosity $L_\gamma$ of young HE pulsars follow the trend $L_\gamma\propto\dot{E}_{\rm rot}^{1/2}$, while MSPs seem to follow the trend $L_\gamma\propto\dot{E}_{\rm rot}$ (see Figure~\ref{fig:medley}; \cite{2PC}), although there is large scatter in the latter case, which may be partly explained by uncertain distances, variations in equation of state (since $\dot{E}_{\rm rot}\propto I$), or different beam and pulsar geometries\footnote{Guillemot \& Tauris~\cite{Guillemot14} find marginal evidence that MSPs that are undetected in $\gamma$ rays may have relatively small viewing angles $\zeta$ with respect to their spin axes, such that their emission beams do not cross Earth's line of sight, or the modulation of the $\gamma$-ray emission may be very weak. In caustic models (Section~\ref{sec:OldTheory}), most of the bright emission occurs near the spin equator, requiring large viewing angles to observe bright, sharp peaks.}~\cite{Guillemot14,Guillemot16,Kalapotharakos17b}. Thus pulsars become increasingly more efficient $\gamma$-ray emitters as they age (converting a larger fraction of $\dot{E}_{\rm rot}$ into $L_\gamma$~\cite{2PC}; cf.\ Figure~\ref{fig:medley}) even though they have smaller $\dot{E}_{\rm rot}$.
  \item GeV power is typically still the dominant component of the multi-frequency spectrum (for all but the youngest Crab-like pulsars).
  \item The HE spectra are quite hard (with an average photon spectral index $\Gamma\sim1.4$) and exhibit spectral cutoff energies\footnote{The spectral cutoff energy $E_{\rm cut}$ seems to scale with some positive power of the $B$-field at the light cylinder $B_{\rm LC}\propto P^{-5/2}\dot{P}^{1/2}$, which may be explained by the fact that the accelerating electric field $E_{||}$ (which is parallel to the local $B$-field) generically scales with $B_{\rm LC}$ in outer-magnetospheric models and in the case of curvature radiation (CR) reaction where the acceleration rate balances this radiative loss rate, one expects $E_{\rm cut}\propto E_{||}^{3/4}$~\cite{2PC}.} $E_{\rm cut}$ in a very narrow band around a few GeV~\cite{2PC} (the soft-$\gamma$-ray pulsars are exceptions, with spectral cutoffs and dominant radiative power occurring in the MeV band~\cite{Kuiper15}). Moreover, this power-law plus sub-exponential spectral shape is characteristic of pulsars and is used to aid in candidate selection for follow-up observations of unidentified \textit{Fermi} sources, in addition to a low variability index that distinguishes them from active galactic nuclei~\cite{Acero13b}. A key result from the \textit{Fermi} Mission was favouring a sub-exponential spectral cutoff over a super-exponential one in the case of the Vela pulsar at a significance of $16\sigma$, indicating that HE emission must come from the outer magnetosphere in order to escape pair production or photon splitting in the intense $B$-field near the stellar surface to reach the observer (e.g.,~\cite{Story14}). Such sub-exponential spectra have now also been seen in other pulsars, possibly indicating a blend of single-particle spectra as the line of sight crosses the pulsar beam~\cite{Vela2,Celik11}. The bulk of pulsars in the Second \textit{Fermi} Pulsar Catalog (2PC), however, favour a simple exponential cutoff while some require a sub-exponential roll-over, but none requires a super-exponential cutoff.
  \item TeV emission from pulsars may be uncommon or intrinsically faint, and therefore rather hard to detect given the current and near-future telescope capabilities: while pulsed photons in the 25~GeV $-$ 1~TeV band have now been detected for the Crab pulsar~\cite{MAGIC_Crab08,MAGIC_Crab11,MAGIC_Crab12,Ansoldi16}, and photons in the 10~GeV $-$ 110~GeV band for the Vela pulsar~\cite{Leung14,DeNaurois15}, McCann's stacked analysis of \textit{Fermi} pulsars~\cite{McCann15} using $\sim4.2$ years of data per pulsar indicates that emission above 50~GeV must be rare for most pulsars, since the average emission per pulsar from a sample of 150 pulsars (excluding the Crab) was limited to lie below $\sim7\%$ of that of the Crab pulsar in the $56-100$~GeV band and below $\sim30\%$ in the $100-177$~GeV band. Burtovoi \textit{et al.}~\cite{Burtovoi17} reanalysed 5 years of \textit{Fermi} LAT data for the 12 pulsars seen to pulse above 25~GeV in the First \textit{Fermi} LAT Catalog of Sources above 10 GeV (1FHL), fitting a power-law spectrum for energies $>10$~GeV and extrapolating this spectrum to energies $>100$~GeV. They predict that the Cherenkov Telescope Array (CTA) may significantly detect up to 8 of these 12 pulsars in 50~h. The Third \textit{Fermi} Catalogue of Hard Sources (3FHL) firmly identifies 53 pulsars above 10~GeV and additionally finds  associations with 6 pulsars known at lower energies~\cite{3FHL}. Interestingly, there are~10 sources in the High Altitude Water Cherenkov (HAWC) Observatory Gamma-Ray Catalog (2HWC) associated with pulsar wind nebulae or supernova remnants~\cite{2HWC} in the energy range between hundreds GeV and tens of TeV; one should therefore distinguish between the ubiquitous TeV pulsar wind nebula emission (e.g.,~\cite{HESS_PWN17}) powered by energetic embedded pulsars (some seen to be pulsing at GeV energies), and the seemingly rare pulsed TeV emission from pulsars\footnote{See also the very recent result on the detection of multi-TeV pulsed photons from the Vela pulsar: https://fskbhe1.puk.ac.za/people/mboett/Texas2017/Djannati.pdf}~(e.g., \cite{Ansoldi16,DeNaurois15}).
  \item The \textit{Fermi} crop of $>200$ pulsar discoveries is diverse\footnote{This is in addition to the detection of modulated emission that originates in high-mass pulsar binaries due to interaction of the pulsar wind with the massive companion's wind and photon field~(e.g.,~\cite{B1259,Xing16}; see also Lyne \textit{et al.}~\cite{Lyne15} for another plausible system of this type).}: there are radio-loud vs.\ radio-faint\footnote{The term ``radio-faint'' is nowadays preferred over ``radio-quiet'', since pulsars might not be absolutely (intrinsically) radio-quiet, but just difficult to detect with current telescopes.} ones, young pulsars vs.\ MSPs, and pulsars in evolving binary systems (redback and black widow systems~\cite{Roberts11}) vs.\ isolated ones~\cite{Smith17}. Surprisingly, MSPs turn out to be a substantial sub-class of HE pulsars, being energetic emitters of GeV emission~\cite{Caraveo14,Grenier15}. Furthermore, blind period searches directly in the $\gamma$-ray data~\cite{SazParkinson2010,Pletsch12}, also using distributed volunteer (crowd) computing~\cite{Clark17}, have made an enormous impact.
  \item Young pulsars occur near the Galactic Plane, while old MSPs are detected at all latitudes, because of the closer distance of these fainter objects and also because old MSPs have had time to evolve to larger scale heights above the Galactic Plane due to their large velocities~\cite{2PC}.
  \item The photon spectral index $\Gamma$ softens with larger $\dot{E}_{\rm rot}$ values~\cite{2PC}, possibly indicating an increase in pair production or the onset of a synchrotron radiation (SR) component in more energetic pulsars.
  \item Radio-quiet $\gamma$-ray MSPs seem to be extremely rare (only two candiates are noted so far~\cite{Acero13a, Kong14} out of a population of $>100$ known ones), which may be attributed to MSPs having very wide $\gamma$-ray and radio beams owing to their relatively compact magnetospheres (since $R_{\rm LC}\sim P$).
 \item Surprising variability was detected in the wind of Crab pulsar~\cite{AGILE_flares,Crab_flares} (i.e., ``Crab flares''), while the pulsed emission remained stable. Another type of variability was found in PSR J2021+4026~\cite{Allafort13}, which exhibited changes in HE flux, light curve morphology, and spectrum coincident with an abrupt step change in spin-down power.
\end{enumerate}

\section{Basic Theoretical Framework}\label{sec:Frame}

\subsection{The Unipolar Inductor -- A Conductor Rotating in a Magnetic Field}
A number of authors have pointed out the similarity between the physics of a unipolar inductor\footnote{Alternative terms include homopolar generator,  unipolar generator, acyclic generator, disk dynamo, or Faraday disc (e.g., \cite{Valone01}).} and a pulsar that is an aligned rotator (having aligned magnetic and spin axes). Consider a conducting disc spinning in a static $B$-field~\cite{Montgomery99}. Electrons in the disc move with a net velocity $\vec{v}=\vec{\Omega}\times\vec{r}$ and experience a Lorentz force $\vec{F}=-e\vec{v}\times\vec{B}/c$ in the surrounding $B$-field. Electrons move toward the axis, leading to a steady configuration in which the total Lorentz force on the electrons vanishes. Similarly, for the aligned rotator in the force-free (FF) limit (plasma-filled, co-rotating magnetosphere and neglecting particle inertia), one finds~\cite{GJ69} 
\beq
\vec{E} + \frac{\vec{\Omega}\times\vec{r}\times\vec{B}}{c} = 0,
\eeq
implying $\vec{E}\cdot\vec{B} = 0$. 
This sets up a potential difference between the axis and rim (or for a pulsar, on the stellar surface between the pole and edge of the PC, which delineates the open $B$-field line region of the magnetosphere, for a pulsar):
\beq
\Delta V = -\int_0^a\vec{E}\cdot d\vec{s} = \frac{\Omega\Phi_{\rm B}}{2\pi} = -\frac{B_0\Omega a^2}{2c},\label{eq:Vdrop}
\eeq
with $\Phi_{\rm B}$ the magnetic flux, $B_0$ the $B$-field, and $a$ the disc radius. There is a component $E_{||}$ of the electric field parallel to the local $B$-field (nearly a radial electric field) associated with this potential drop, which pulls primary charges from the stellar surface and eventually fills the magnetosphere with plasma via ensuing HE emission and a cascade of secondary $e^+/e^{-}$ pair production (Section~\ref{sec:OldTheory}), creating an FF magnetosphere. Using Gauss' law as well as the electric field that occurs in such an FF magnetosphere where $\vec{E}\cdot\vec{B} = 0$, we find the so-called Goldreich-Julian charge density~\cite{GJ69}
\beq
\rho_{\rm GJ}= \frac{\vec{\nabla}\cdot\vec{E}}{4\pi}\approx\frac{\vec{\Omega}\cdot\vec{B}}{2\pi c}.\label{eq:rhoGJ}
\eeq
The radius where the corotation speed $|\vec{v}_{\rm rot}|=|\vec{\Omega}\times\vec{r}|=c$, is
\beq
R_{\rm LC} = \frac{c}{\Omega}\propto P.
\eeq
This is the so-called ``light cylinder radius'' that sets the typical spatial scale for the pulsar magnetosphere. The last open field line tangent to the light cylinder defines the PC, the rim of which lies at a polar angle (in the aligned case)
\beq
\Theta_{\rm PC} = \left[\sin^{-1}\left(\frac{\Omega R}{c}\right)\right]^{1/2}\approx\left(\frac{\Omega R}{c}\right)^{1/2},
\eeq
with $R$ the stellar radius.

\subsection{The Braking Model -- Inference of Fundamental Quantities}
Pulsars are born as remnants of supernova explosions following the gravitational collapse of a massive star~\cite{BaadeZwicky34}. For a stellar core that rotates more or less rigidly and assuming that the angular momentum is conserved during collapse, the final angular velocity will be
\beq
\Omega_f \sim \Omega_i\left(\frac{R_i}{R_f}\right)^2,
\eeq
with $R$ and $\Omega=2\pi/P$ the radius and angular velocity, and ``i'' and ``f'' indicating the initial and final values. Inserting typical values of $R_i \sim 10^{11}$ cm and $R_f\sim 10^6$ cm into the above equation yields an increase in angular velocity by a factor of $\sim 10^{10}$ and rotational periods in the millisecond to second range. If the stellar interior is fully conductive, magnetic flux ($\Phi_B \equiv \oint \vec{B} \cdot d\vec{a} \sim B_iR_i^2$) will also be conserved during collapse, implying\beq
B_f \sim B_i\left(\frac{R_i}{R_f}\right)^2.
\eeq
This relation yields typical surface $B$-fields of $B_0\sim 10^{12}$ G. 

Rotational energy is the reservoir that is tapped and converted into electromagnetic (fields, pulsed emission) and particle (pulsar wind) energy. An isolated neutron star will thus ``spin down'' and rotate slower (i.e., $\dot{P}>0$). An estimate for the surface polar $B$-field strength may be obtained by equating the rate of slowing down and the magnetic dipole radiation loss rate\footnote{Beskin \textit{et al.}~\cite{Beskin83} argued that longitudinal magnetospheric currents determine the pulsar spin-down. Recently, Beskin \textit{et al.}~\cite{Beskin17} argued that pulsar slow-down or braking is (additionally and possibly predominantly) due to the separatrix currents that circulate in the pulsar magnetosphere and do not flow out as part of the pulsar wind. Spitkovsky~\cite{Spitkovsky2006} noted that the spin-down due to Poynting flux leaving an FF (plasma-filled) magnetosphere is similar to the vacuum case of magneto-dipole losses, but with the $\sin^2\alpha$ term replaced by $1+\sin^2\alpha$. In reality, pulsed emission also taps energy from the pulsar's rotational energy (i.e., $L_\gamma$ must be some fraction of $\dot{E}_{\rm rot}$), while a pulsar wind beyond the termination shock blows a pulsar wind nebula, following the conversion of much of the electromagnetic energy to particle acceleration. This expression for loss rate also neglects losses due to gravitational waves, since the neutron star is expected to be close to spherical in shape, although some have taken distortions or precession that may produce gravitational wave emission into account (e.g.,~\cite{Pandharipande76,Bonazzola96,Zimmerman80}).} for a star in vacuum~\cite{Ostriker69}:
\beq
\dot{E}_{\rm rot}\equiv\frac{d}{dt}\left(\frac{1}{2}I\Omega^2\right) = I\Omega\dot{\Omega} = -\frac{4\pi^2I\dot{P}}{P^3}=L_{\rm md}= -\frac{2}{3c^3}\mu^2\sin^2\alpha\,\Omega^4,\label{eq:dipole_loss}
\eeq
with $L_{\rm md}$ the loss rate due to magneto-dipole radiation, $I\sim MR^2$, $\mu \equiv B_0R^3/2$ the magnetic moment, $B_0$ the $B$-field at the pole, $\alpha$ the angle between the magnetic and spin axes, and $c$ the speed of light. Thus, if one assumes that $\mu\sin\alpha\approx {\rm const.}$, the general ``braking'' or ``spin-down'' law may be written as
\beq
\dot{\Omega} \propto -\Omega^n,\label{eq:law}
\eeq
with $n$ the braking index that may be obtained by differentiating the above equation with respect to time (for constant $n\neq1$):
\beq
n = \frac{\ddot{\Omega}\Omega}{\dot{\Omega}^2} = 2 - \frac{P\ddot{P}}{\dot{P}^2},
\eeq
with $\ddot{\Omega}$ and $\ddot{P}$ the second derivate of the angular speed and period, and\footnote{It is clear that setting $n=3$ is a simplification: a recent accumulation~\cite{Archibald16} of measured values for $n$ ranges from 0.9 to 3.15; possible explanations for the deviation of $n$ from 3 may include angular momentum loss due to the pulsar wind, effects of a quadrupole moment of the $B$-field on the braking evolution, and glitches.} $n=3$ for a dipolar $B$-field.
By inserting typical values of $I \sim 10^{45}$ g cm$^2$, $R \sim 10^6$ cm and $\alpha \sim 90^{\circ}$ into Eq.~(\ref{eq:dipole_loss}) one obtains an estimate for the surface $B$-field at the pole:
\beq
B_0 \sim 6 \times 10^{19}P^{1/2}\dot{P}^{1/2}~{\rm G}.\label{eq:B_derived}
\eeq
By assuming that the $B$-field does not decay over the Myr timescales for young HE pulsars\footnote{Some population synthesis models do include $B$-field decay, however; cf.~\cite{Han97,Igoshev15,Tauris01}.}, and a constant inclination angle $\alpha$, one finds that~\cite{Venter07} 
\beq
\dot{P}P^{n-2} = K,
\eeq
with $K$ a constant. Integration of the above leads to (e.g., \cite{Venter15})
\beq
P^{n-1} = P_0^{n-1}(n-1)K,
\eeq
with $P_0$ the birth period. Assuming the $B$-field is dipolar, one may adopt an $r^{-3}$ dependence and calculate the poloidal field at the light cylinder\footnote{Interestingly, Michel~\cite{Michel91} notes that $L_{\rm md}$ (Eq.~\ref{eq:dipole_loss}) may be estimated by multiplying the energy density contained in $B_{\rm LC}$ by the area $4\pi R^2_{\rm LC}$ and $c$.} (the toroidal field starts to dominate beyond $R_{\rm LC}$ and has an $r^{-1}$ dependence):
\beq
B_{\rm LC}= B_0\left(\frac{R}{R_{\rm LC}}\right)^3\propto P^{-5/2}\dot{P}^{1/2}.\label{eq:BLC}
\eeq
By assuming that $\mu_{\bot} \equiv \mu\sin\alpha$ remains roughly constant, and so does\footnote{Some authors have investigated scenarios in which this is not the case~\cite{Aris17,Tauris01}.} $n$, the characteristic or ``rotational'' age $\tau_{\rm c}$ can be derived upon integration of Equation~(\ref{eq:law}) and substitution of $\Omega^{n-1} = \dot{\Omega}/(k\Omega)$, with $k$ a constant:
\beq
\tau_{\rm c} = -\frac{\Omega}{(n-1)\dot{\Omega}}\left[1-\left(\frac{\Omega}{\Omega_0}\right)^{n-1}\right] \approx -\frac{\Omega}{(n-1)\dot{\Omega}} \equiv \frac{P}{(n-1)\dot{P}} \label{eq:age},
\eeq
when $\Omega_0 \gg \Omega$.
\begin{figure}[t]
  \begin{center}
  \includegraphics[width=12cm]{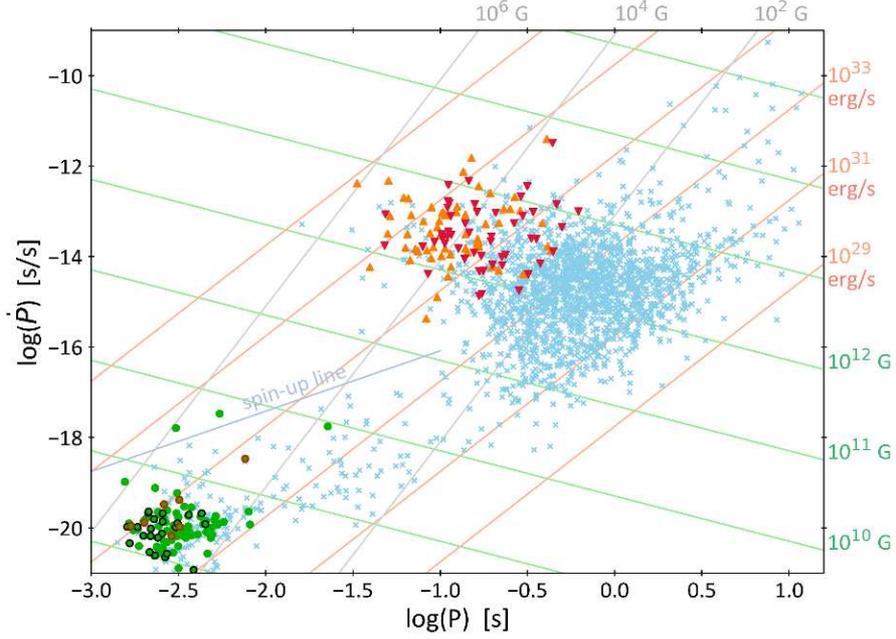}
  \caption{Updated plot~\cite{Grenier15} of pulsar period ($P$) vs.\ period time derivative ($\dot{P}$) of $\gamma$-ray and radio pulsars, with 53 radio-loud and
$\gamma$-loud young pulsars (orange upward triangles), 37 radio-faint and $\gamma$-loud young pulsars (red downward triangles), 71 radio-loud and
$\gamma$-loud MSPs (green filled circles, circled in black and red when in black-widow and redback systems, respectively), and
2~256 other radio pulsars (light blue crosses). %Recently discovered MSPs, with no $\dot{P}$ measurement yet, are plotted as squares at $\dot{P}$ near $10^{-21}$. 
Lines of constant spin-down power (brown; (Eq.~[\ref{eq:dipole_loss}])) and polar $B$-field strength (green; Eq.~[\ref{eq:B_derived}]) are given for a magnetic
dipole in vacuum and a stellar moment of inertia of $1.4 \times 10^{45}$ g cm$^{-2}$ applicable to a 1.4 solar mass neutron star with a 12~km
radius. Lines of constant $B$-field strength at the light cylinder (Eq.~[\ref{eq:BLC}]) radius are shown in grey. The bluish-grey line marks the spin-up rate expected from mass transfer at the Eddington rate from a stellar companion in a binary system.}
  \label{fig:PPdot}
  \end{center}
\end{figure}
The PC voltage may be written as (by substituting $a=R\sin\Theta_{\rm PC}$ into Eq.~[\ref{eq:Vdrop}])
\beq
-\Delta V_{\rm PC} = \frac{B_0\Omega^2R^3}{2c^2} \sim|L_{\rm md}|^{1/2}.
\eeq
The Goldreich-Julian current is (Eq.~[\ref{eq:rhoGJ}])
\beq
I_{\rm GJ}\sim2\rho_{\rm GJ}cA\sim|L_{\rm md}|^{1/2},
\eeq
with $A=\pi R^2\sin^2\Theta_{\rm PC}$ the area of one PC.
The total electromagnetic power is thus
\beq
L=\Delta V_{\rm PC}I_{\rm GJ}\sim|L_{\rm md}|.
\eeq
If one accepts a constant $\Delta V$ as a threshold condition\footnote{The $\Delta V$ threshold is a direct consequence of the self-controlled feedback between $E_{||}$ and the acceleration length required for the primary particles to reach $\gamma$-ray emitting energies, and the subsequent screening of $E_{||}$ by the secondary pairs. Thus, $\Delta V$ remains nearly constant for different intial conditions for $E_{||}$.} for pair production in young pulsars~\cite{Harding81}, one expects the $\gamma$-ray luminosity to behave as (assuming $\dot{E}_{\rm rot}\sim L_{\rm md}$; Eq.~[\ref{eq:dipole_loss}]) 
\beq
L_\gamma\sim \Delta V_0I_{\rm GJ}\sim\dot{E}_{\rm rot}^{1/2},
\eeq
since $\Delta V = \Delta V_0 = {\rm constant}$ in this case. On the other hand, if older pulsars have pair-starved magnetospheres, their $\gamma$-ray luminosity may behave as
\beq
L_\gamma\sim \Delta V_{\rm PC}I_{\rm GJ}\sim\dot{E}_{\rm rot}.
\eeq
Using the above expressions for $\dot{E}_{\rm rot}$, $B_{\rm LC}$, $B_0$, and $\tau_{\rm c}$, a $\dot{P}P$-diagram may be constructed to categorise the various pulsar species we observe (Figure~\ref{fig:PPdot}). Although many basic properties may be succinctly summarised in this plot, there are surely a number of simplified assumptions that need to be revisited to obtain a more realistic summary plot for the population of pulsars (e.g., effects of considering $\dot{B}$, $\dot{\alpha}$, equation of state, pulsar winds, distribution of birth periods and $B$-fields, etc.).

\section{Standard Emission Models -- Explaining Spectra and Light Curves}\label{sec:OldTheory}
The aligned-rotator model of Goldreich and Julian~\cite{GJ69} provided an ``existence proof'' for a plasma-filled magnetosphere: the rotationally-induced electric field vastly dominates gravity (and particle inertia) near the stellar surface, ripping charges from the crust and accelerating these primary charges along the nearby $B$-field lines. In PC~\cite{DH96} and slot gap (SG)~\cite{Arons83} models, the primary particles emit CR as they are constrained to move along curved field lines. The HE photons undergo magnetic (one-photon) pair creation and this leads to a cascade of electron-positron pairs that fill the surrounding magnetosphere and screen the electric field $E_{||}$. However, there remain regions (just above the stellar surface in PC models, before the pair formation front develops at a fraction of $R$ in altitude; and along the last closed field lines in SG models where $E_{||}$ vanishes and the pair formation mean free path becomes infinite) where the plasma is not dense enough to shield $E_{||}$, and particle acceleration can take place. In OG models~\cite{CHR86,R96}, particle outflow above the null-charge surface (where $\vec{\Omega}\perp\vec{B}$ and $\rho_{\rm GJ}=0$) creates gaps where acceleration and two-photon pair production may take place. Pair-starved polar cap (PSPC) models have been studied in the context of suppressed pair production in older pulsars~\cite{HUM05}. Annular and core gap models~\cite{Qiao04,Du11} invoke gaps between critical field lines (lines that intersect the null-charge surface at the light cylinder) and last-closed field lines.

Early on, fundamental electrodynamical questions emerged: How and where is the current closed so that the outflow of particles is sustainable (i.e., what is the global current flow pattern)? What is the role of pair formation? What is the injection rate of plasma from the stellar surface? Where do acceleration gaps develop (where $E_{||}$ is not fully screened) and how are they sustained? Where does acceleration of particles to relativistic energies take place? What is the emission mechanism for each of the multi-frequency components we observe?

In the interum period leading up to the development of global emission models, geometric two-pole caustic (TPC)~\cite{Bai10,DR03,Dyks04} and OG~\cite{Venter09,Watters09} light curve models were used to constrain emission gap and pulsar geometries (inclination and observer angles $\alpha$ and $\zeta$). Such geometric models do not contain any knowledge of the $E_{||}$ distribution\footnote{Thus one can not calculate a spectrum or energy-dependent light curves from such models.}, but rather assume a constant emissivity per unit length in the corotating frame along certain $B$-field lines, with photons being emitted tangentially to the local field lines in the corotating frame. Aberration plus time-of-flight delays are included, leading to photons bunching in phase to form so-called caustics of bright emission. These caustics result in sharp peaks as the asymmetric beam sweeps past the observer. Although these models had reasonable success in reproducing HE light curves~\cite{Johnson14, Pierbattista15,Venter09,Venter12}, they pointed to the fact that a more general model is needed of which the various geometric models may be particular incarnations~\cite{Venter14}.

In addition to the local gap models interior to the light cylinder, work has also been done on ``striped-wind'' models, where dissipation takes place in the current sheet that forms near the equator beyond the light cylinder (e.g., \cite{Petri11,Petri12}). The notion of the current sheet emerges in the context of FF $B$-field models. In contrast to the rotating vacuum dipole solution obtained by Deutsch~\cite{Deutsch55}, which has been used in several pulsar light curve models as a first approximation~(e.g., \cite{Dyks04,Venter09}), the FF solution assumes that there is dense enough plasma everywhere so that $E_{||}$ may be screened throughout the magnetosphere. This leads to the ``pulsar equation'' that has been solved for the aligned-rotator case~\cite{CFK99}. The FF field has also been obtained for the oblique~\cite{Spitkovsky2006} case, and additionally using full MHD~\cite{Komissarov07, Tchekhovskoy13}. 

However, both vacuum or plasma-filled (FF) pulsar magnetospheres can only be extreme approximations to reality, since the first possesses no charges to radiate the pulsed emission we observe, while the latter permits no electric fields $E_{||}$ that may accelerate charges to high enough energies to radiate HE emission. Dissipative magnetosphere MHD solutions~\cite{Kalapotharakos12, Kalapotharakos14, Li12} seek to obtain more realistic solutions by including a macroscopic conductivity $\sigma$ as a free parameter, and therefore allowing charges, currents, and acceleration to occur in the pulsar magnetosphere. The question of how $\sigma$ comes about must be closely linked to how injection and pair formation rates differ in different regions in the magnetosphere. Particle-in-cell (PIC) codes study such microphysical questions, but are subject to computational limits as well as particular assumptions on initial magnetospheric configuration. See Venter~\cite{Venter16_HEASA} and references therein for a more detailed overview of the above models.

\section{New Theoretical Developments -- How Do We Refine Our Models?}\label{sec:Theory}
\subsection{Dissipative Magnetic Fields}
Dissipative models have been developed~\cite{Kalapotharakos12, Kalapotharakos14,Li12} allowing solutions that transition from the vacuum to FF case (from zero to formally infinite $\sigma$). Kalapotharakos \textit{et al.}~\cite{Kalapotharakos14} attempted to model the observed inverse correlation between the $\gamma$-ray peak separation in phase, $\Delta$, and the radio-to-$\gamma$ phase lag, $\delta$, as this would potentially constrain the $\sigma$-distribution. They found that this so-called $\Delta-\delta$ trend could only be reproduced for a spatially-dependent macroscopic $\sigma$: FF conditions should exist interior to the light cylinder, and a large but finite $\sigma$ outside. These models are referred to as FIDO models -- FF inside, Dissipative Outside. Brambilla \textit{et al.}~\cite{Brambilla15} next calculated phase-averaged and phase-resolved spectra predicted by the FIDO model for a few very luminous pulsars assuming CR. They found that the $\gamma$-ray flux and spectral cutoff energy $E_{\rm cut}$ generally increased for larger observer angles $\zeta$, and decreased for an increase in $\sigma$ (for a fixed obliquity $\alpha$). These quantities also increased with decreasing period $P$ and increasing surface $B$-field $B_0$. The FIDO model furthermore predicted that $E_{\rm cut}$ of phase-resolved spectra may typically increase near the phase of the second light curve peak (in the case of $\sim75\%$ of the predicted light curves), but opposite behaviour was also noted. Importantly, they found a tentative correlation between $\sigma$ and $\dot{E}_{\rm rot}$ as well as an anti-correlation between $\sigma$ and age $\tau_{\rm c}$. These trends are expected if one identifies higher $\sigma$ with more efficient screening of $E_{||}$ by pairs (which possibly happens in younger, more energetic pulsars).

\begin{figure}[t]
  \begin{center}
  \includegraphics[width=15cm]{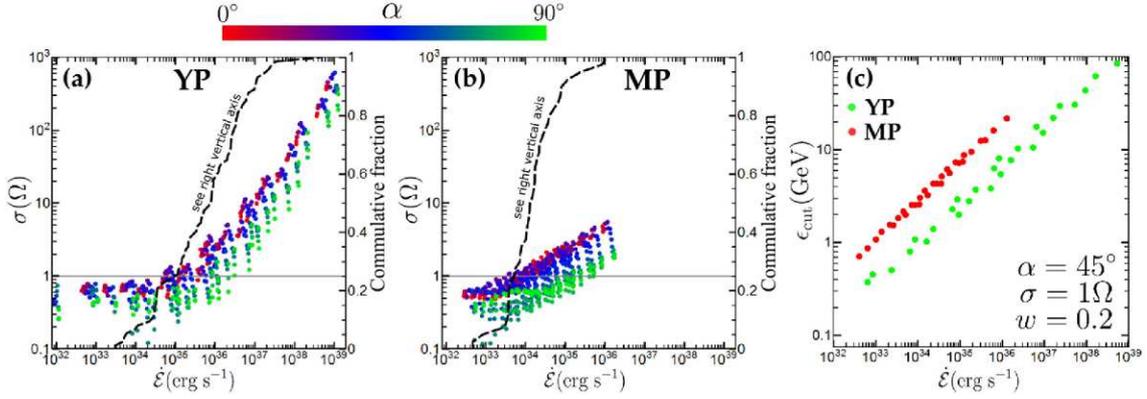}
  \caption{\textit{Panel~(a):} The optimal $\sigma$ values found by Kalapotharakos \textit{et al.}~\cite{Kalapotharakos17a} for young pulsars, for a gap width of $w=0.1$, that reproduce the spectral cutoffs measured by \textit{Fermi} LAT. The values of $\sigma$ below the grey lines were derived by extrapolation, and require larger $w$-values. The cumulative fraction (distribution with $\dot{E}_{\rm rot}$) of the corresponding \textit{Fermi} pulsars is indicated by a dashed line, as noted on the right vertical axis. \textit{Panel~(b):} The same as \textit{Panel~(a)}, but for MSPs. \textit{Panel~(c):} The predicted spectral cutoffs corresponding to the parameter values indicated in the legend, for both young pulsars and MSPs (labelled YP and MP).}
  \label{fig:Kala}
  \end{center}
\end{figure}

The work of Kalapotharakos \textit{et al.}~\cite{Kalapotharakos14} indicated that uniform, high-$\sigma$ models lead to dissipation happening predominantly near the current sheet. This motivated Kalapotharakos \textit{et al.}~\cite{Kalapotharakos17a} to refine the FIDO model by applying a low $\sigma$ only in a narrow region near the current sheet outside the light cylinder, near the open-field-line boundary (PC rim), while still keeping a high $\sigma$ inside the light cylinder. This preserved the FF structure, especially for low $\alpha$. They demonstrated the inverse trend between the measured HE cutoff energy $E_{\rm cut}$ and model $\sigma$ at different $\dot{E}_{\rm rot}$ (e.g., for high $\sigma$, $E_{||}\propto\sigma^{-1}$, and in turn $E_{\rm cut}\sim E^{3/4}_{||}$ in the CR radiation-reaction limit). Invoking CR and assuming that the radiation-reaction limit is reached, they inferred $E_{||}$ implied by the measured value of $E_{\rm cut}$ by \textit{Fermi} for a population of pulsars. They could show that this $E_{||}$ decreases with $\dot{E}_{\rm rot}$ but saturates at low $\dot{E}_{\rm rot}$. They next constructed spectra for typical values of $\sigma$, $\alpha$, and $\dot{E}_{\rm rot}$ and fit these to data, thereby inferring the optimal $\sigma_{\rm opt}$ for each $\dot{E}_{\rm rot}$ (Figure~\ref{fig:Kala}). The optimal $\sigma_{\rm opt}$ decreased with $\alpha$ for a fixed $\dot{E}_{\rm rot}$. From this followed the trends $\sigma_{\rm opt}\propto\dot{E}_{\rm rot}$ for young pulsars and $\sigma_{\rm opt}\propto\dot{E}_{\rm rot}^{1/3}$ for MSPs. This implies higher pair creation rates (and $\sigma$) for more energetic pulsars. For lower values of $\dot{E}_{\rm rot}$, a wider gap was needed to fit the data. Thus, a larger dissipation region was needed to increase $E_{||}$ (decrease $\sigma$) and thus increase the predicted $E_{\rm cut}$. Lastly, by comparing model predictions to an obervationally-inferred $L_\gamma$ vs.\ $\dot{E}_{\rm rot}$ plot, they found that their model did not yield large enough particle multiplicities\footnote{The number of secondaries spawned via pair production of radiation produced by primary particles; high multiplicities imply large currents.} to account for the measured $L_\gamma$ at very high $\dot{E}_{\rm rot}$ and $\alpha$ values. This model thus provides a tantalising macroscopic description of pulsars that may guide kinetic codes attempting to uncover the microphysics that support the required macroscopic charge and current densities. 

The conductivity in the macroscopic dissipative models is supplied by the electron-positron pair plasma at a microscopic level. Where and how this pair plasma originates is currently not completely understood. Important constraints on the location of pair plasma production in the pulsar magnetosphere as well as its spectrum will come from observations in the 100~keV to 10~MeV band. Radiation models suggest that non-thermal emission in this band comes from SR of electron-positron pairs, produced either at the PCs \cite{Harding15} or the OGs \cite{Takata08}. The shape of the spectral energy distribution reflects the pair spectrum and its peak can determine where the pairs are produced \cite{DeAngelis17,Harding17}. Observations of pulsars whose non-thermal emission peaks in the MeV band with proposed telescopes with superior sensitivity such as AMEGO \cite{Moiseev2017} and e-ASTROGAM \cite{DeAngelis17} are anticipated in the next fifteen years.

\subsection{PIC Codes}
\begin{figure}[t]
  \begin{center}
  \includegraphics[width=13cm]{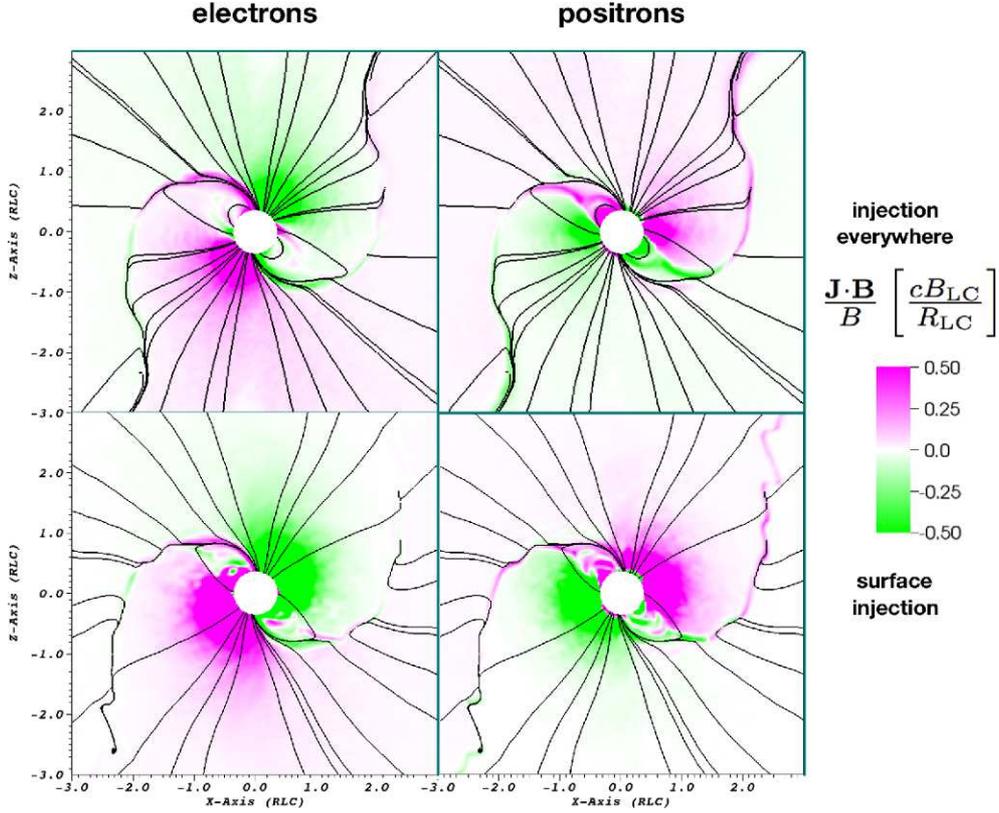}
  \caption{The electron and positron components of the current density for magnetospheres close to being FF, as predicted by the PIC model of Brambilla \textit{et al.}~\cite{Brambilla18}.  One can see that the electrons and positrons both flowed out in the PC regions. The labels distinguish the cases where pair injection took place only at the surface vs.\ everywhere in the magnetosphere.}
  \label{fig:Bram}
  \end{center}
\end{figure}

The application of kinetic PIC codes to pulsar magnetospheres marks a mini-revolution in theoretical studies of neutron stars. This technique can model the magnetosphere from first principles, in contrast to the approaches described above. It is important to resolve both the temporal and spatial scales of the problem (plasma frequency and skin depth) to avoid numerical instability and numerical plasma heating~\cite{Brambilla18}. Correcting for the effect of too low a $B$-field\footnote{A low $B$-field and low voltage are required to resolve the skin depth for the highest-energy particles.} on the radiative properties is also important to fully capture the emission physics~\cite{Kalapotharakos17b}. Previous works~\cite{Belyaev15a,Belyaev15b,Chen14,Cerutti15,Cerutti16a,Cerutti16b,Cerutti17b,Philippov14,Philippov15a,Philippov15b,Kalapotharakos17b} focused on dealing self-consistently with the pulsar electrodynamics including global current closure, the contribution of charges of different sign to the current, dissipative processes, electromagnetic emission, and the effects of pair production and general relativity (see Venter~\cite{Venter16_HEASA} for a more detailed summary). Several aspects, including the importance of (spatial) particle injection properties (which was found to critically depend on general relativity), as well as a renewed focus on the current sheet and Y-point\footnote{This is a region of merging field lines close to the light cylinder, where the inner magnetospheric lines transition to a equatorial current sheet~\cite{Timokhin06}.} as important dissipative regions (including the study of plasma instabilities and magnetic reconnection) came to the fore. Here, we will only describe two studies that represent some of the most recent work in this area.

Brambilla \textit{et al.}~\cite{Brambilla18} studied current composition and flow using a new PIC code \cite{Kalapotharakos17b}, focusing on the dependence of magnetospheric properties on particle injection rate. This study is more realistic than some previous works given the larger particle injection rates that could be attained. As in prior studies (e.g., \cite{Belyaev15b,Cerutti15}), they obtained a transition of the magnetospheric solutions from vacuum to FF (as seen in MHD models), invoking two scenarios: particle injection from the stellar surface and injection everywhere in the magnetosphere. They found the highest dissipation (larger than in the FF case) for intermediate injection rates. As the injection rate was increased (i.e., equivalent to a macroscopic $\sigma$ being increased), $E_{||}$ was gradually (but never completely) screened and the FF current structure was attained. Although these two injection scenarios provide similar field structure and current density distributions, they differ in particle density distribution in the sense that higher multiplicities were reached at the neutron star surface in the surface-injection scenario. Furthermore, charges in this case rearranged themselves in such a way that charges of one sign may contribute significantly to the current density in regions of opposite charge density, with electron and positron currents almost cancelling each other when $E_{||}$ is nearly screened. Availability of pairs (or not) thus significantly impacts current flow. By studying particle trajectories, they could probe some details of the current composition, e.g., they found that electrons and positrons both flowed out in the PC regions (which may inhibit two-photon pair production; cf.\ Figure~\ref{fig:Bram}), lower-energy electrons returned to the stellar surface by crossing $B$-field lines close to the return current sheet inside the light cylinder (making them ideal candidates for emitting SR in the MeV range), some electrons were semi-trapped near the Y-point before returning to the star, and positrons flowed out on the separatrix where they were accelerated close to the Y-point and collected at the current sheet outside the light cylinder. Such energetic particles flowing out along the current sheet correspond well to the FIDO model assumption~\cite{Kalapotharakos17a} that invokes dissipation regions only near the current sheet beyond the light cylinder. This model generally provides a good description of the \textit{Fermi} pulsar phenomenology. Thus, the latest PIC simulations are now elucidating and justifying the FIDO macroscopic assumptions and electrodynamical (or spatial accelerator) constraints derived from the HE data when assuming CR from positrons in the current sheet~\cite{Kalapotharakos17b}. Future studies should keep reaching for higher particle energies to simulate the radiative physics more realistically. Alternative assumptions of pair production should lead to distinct observational characteristics, which may be probed by future X-ray missions.

\begin{figure}[t]
  \begin{center}
  \includegraphics[width=12cm]{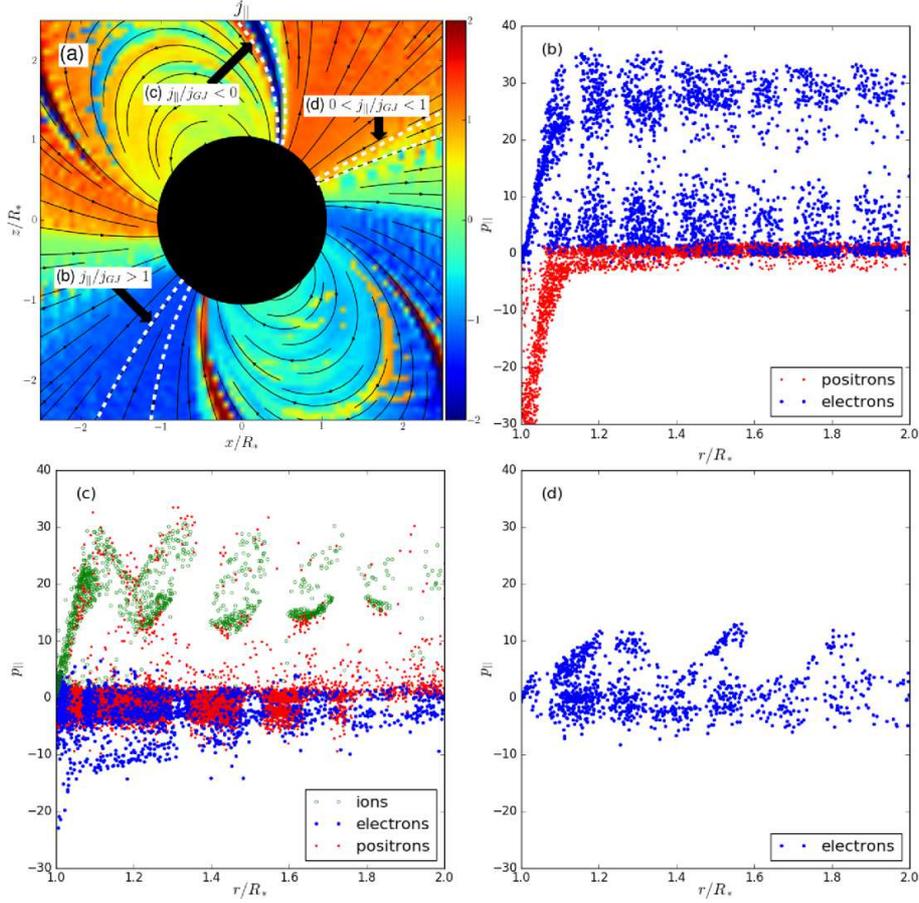}
  \caption{\textit{Panel~(a):} Current distribution in a slice through the $\bm{\mu}-\bm{\Omega}$ plane for a rotator with $\alpha=60^\circ$. The black lines indicate  poloidal $B$-field lines while the colour shows the current component parallel to the $B$-field, normalised by $\Omega B/2\pi$. The white dashed lines indicate boundaries to regions with qualitatively different current flow (and thus pair formation) properties. \textit{Panel~(b):} region with outflowing super-GJ current, $j_{||}/j_{\rm GJ}> 1$; \textit{Panel~(c):} return current, $j_{||}/j_{\rm GJ} < 0$; \textit{Panel~(d):} outflowing sub-GJ current, where $0 < j_{||}/j_{\rm GJ} < 1$.  Ions, electrons, and positrons are indicated by green, blue, and red dots.  Particle momentum is normalised by $m_e c$ (with $m_e$ the electron mass) and the distance from the star, $r$ is measured along the $B$-field lines and normalised by $R$.  One can see that electrons are accelerating outward and positrons inward above the PC in \textit{Panel~(b)}.}
  \label{fig:Phil}
  \end{center}
\end{figure}

Philippov \textit{et al.}~\cite{Philippov17} performed HE emission modelling in their PIC simulations of oblique rotators, including one-photon and two-photon pair production, frame-dragging effects (which substantially increase the number of open field lines that can sustain pair production), electron and ion extraction from the stellar surface (with the ions carrying a significant part of the energy in the outflowing energy flux), and HE photon emission assuming SR. The resulting  particle density and particle energy flux of the pulsar wind varied significantly with latitude. Non-stationary pair creation thus occurred above the PC and also in the return current layer and current sheet. These detailed simulations are also providing important hints as to how the different species of particles make up the global current flow patterns (Figure~\ref{fig:Phil}). Above nearly the whole PC, electrons were accelerated outwards and positrons inward\footnote{This is different from what Brambilla \textit{et al.}~\cite{Brambilla18} found, given the different respective injection assumptions: surface pair injection vs.\ pair formation over the full magnetosphere here (Philippov \textit{et al.}~\cite{Philippov17} assume pair production wherever the particles reach a threshold energy).}, while pair cascades screen $E_{||}$ and the newly-formed pair cloud escaped from this region. The electrons orbited the current sheet where they contributed to the HE photon flux. Some escaped from the pulsar magnetosphere while others returned to the star. Conversely, in the return current layer, $E_{||}$ extracted ions from the surface and both ions and positrons accelerated outward, while there were electrons returning from the Y-point, initiating further pair discharges. Ions and charge-separated plasma, which formed from photons emitted towards the star, dominated the closed-field-line region. The density gap outside the current sheet found in previous studies remained present as a region of suppressed particle density. Philippov \textit{et al.}~\cite{Philippov17} found that SR, produced by particles (mostly positrons) that are accelerated by relativistic magnetic reconnection in the current sheet and close to the Y-point, dominated the $\gamma$-ray waveband emission. Since the $B$-field approaches zero in the current sheet, SR losses were significantly inhibited and the positrons could thus attain very high energies there. The preliminary light curves predictions exhibited the double-peaked structure seen in many \textit{Fermi} pulsars. The caustics in the sky maps traced the current sheet, the projection of which was sinusoidal for low $\alpha$, but the caustics became disconnected for larger $\alpha$ and the contribution from the separatrix layer decreased in this case. The Poynting flux was strongly concentrated near the equator, and the plasma energy flux and density distribution were very non-uniform. These calculations thus yielded the pulsar wind structure at its base and may be useful in solving the ``$\sigma$-problem''\footnote{This is the problem of how electromagnetic energy density is converted into particle energy density on short spatial scales at the termination shock. This $\sigma$ (ratio of electromagnetic to particle energy density) is not to be confused with the macroscopic conductivity used elsewhere.}. Particle acceleration, emission spectral cutoff $E_{\rm cut}$ (which depends on $B_{\rm LC}$), and conversion efficiency of $\dot{E}_{\rm rot}$ to HE emission were found to strongly depend both on obliquity $\alpha$ and the level of pair production in the current sheet (the radiative efficiency decreasing with increasing $\alpha$ and increasing efficiency of pair production in the current sheet). This may help explain the scatter in the $L_\gamma$ vs.\ $\dot{E}_{\rm rot}$ plot.

From these two recent papers, we can see that PIC codes are starting to address very interesting, detailed questions about current flow and emission properties of pulsars, while also raising new questions pertaining to the specific properties of pair production. The latter is fundamentally linked to the particle energetics and radiative output of a pulsar.

\subsection{Other ideas: Multipolar Fields and Polarisation}
A number of authors have pointed out the need for offset-dipole or multipolar $B$-fields beyond the usual assumption of a dipolar rotator (e.g.,~\cite{Arons79,Asseo02,Chen93,RS75}), as also motivated by observations~\cite{Bogdanov2007,Bogdanov2009,Gil2001,Kuzmin86}. Harding \& Muslimov~\cite{HM11a,HM11b} found that introduction of a modest azimuthally asymmetric distortion in the $B$-field (the ``offset-PC model'', cf.~\cite{Barnard16}), which may be due to $B$-field-line sweepback near the light cylinder or non-symmetric currents within the star, can significantly increase $E_{||}$ on one side of the PC. This, combined with a smaller $B$- field line radius of curvature, leads to larger pair multiplicity and a significant extension of pair spectra to lower energies, thus providing a mechanism for pair creation even in (old) pulsars that have previously been thought to be pair-starved. 

P\'etri studied offset-dipole $B$-fields in vacuum in a series of papers. He presented analytical solutions in closed form in flat vaccum spacetime for the retarded point quadrupole, hexapole, and octopole as generalisations of the retarded point dipole, emphasising the effect of $B$-field topology on emitted Poynting flux, braking index, PC geometry, and caustic beam structure. He cautioned that the polar $B$-field strength inferred from observations assuming a dipolar field may be in error in the presence of significant multipolar components~\cite{Petri15}. He next provided analytical solutions for a displaced dipolar field, and computed the $\dot{E}_{\rm rot}$ and the torque exerted on the pulsar's crust, pointing out that HE light curve and polarisation modelling may help constrain the magnetic topology~\cite{Petri16} (see also Kundu \& P\'etri~\cite{Kundu17}). In a dedicated paper, polarisation properties in an off-centre dipole field were studied by extending the well-known rotating vector model to a form appropriate for this topology, called the decentred rotating vector model (DRVM)~\cite{Petri17a}. Finally, P\'etri~\cite{Petri17b} generalised multipolar field expressions to include the effect of strong gravity by computing general-relativistic extensions of the Deutsch solution~\cite{Deutsch55}, including spacetime curvature and frame-dragging effects (both numerically and analytically, but approximately in the latter case).

Gralla \textit{et al.}~\cite{Gralla16} implemented a general analytical procedure for studying an arbitrary axisymmetric FF $B$-field of a slowly rotating star (aligned rotator) including general-relativistic effects, with the dipolar $B$-field component dominating far from the star. They could confirm that conditions conducive to pair production exist above the PC (compatible with recent PIC simulations~\cite{Philippov15b}), even for such non-dipolar fields, and showed that the dipolar component is $\sim60\%$ larger than the canonical value. The location and shape of the PCs are, however, modified dramatically upon inclusion of multipolar field components (becoming offset and even filling an annular region). This may have implications for MSP radio beams as well as phase lags between radio and $\gamma$-ray beams. Gralla \textit{et al.}~\cite{Gralla17} next studied oblique rotators, with the $B$-field being symmetric about an axis other than the rotation axis, including general-relativistic effects and multipole components and focusing on the near-field charge and current flow. Using the method of matched asymptotic expansions\footnote{Solving and matching expressions that are valid near and far from the stellar surface: near the star the equations describe a static vacuum $B$-field in the Schwarzschild spacetime, while far away the FF magnetosphere of a rotating point dipole in flat spacetime is solved numerically.}, they derived a general analytic formula for the PC shape and charge-current distribution as a function of the stellar mass, radius, rotation rate, moment of inertia, and $B$-field. For combined dipole and quadrupole components, thin annular PCs were again obtained. These results may be important for PC heating and resulting X-ray thermal emission calculations, as well as neutron star mass and radius measurements by, e.g., the \textit{NICER} Mission~\cite{Arzoumanian14}, and for pair production physics.

\begin{figure}[t]
  \begin{center}
  \includegraphics[width=14cm]{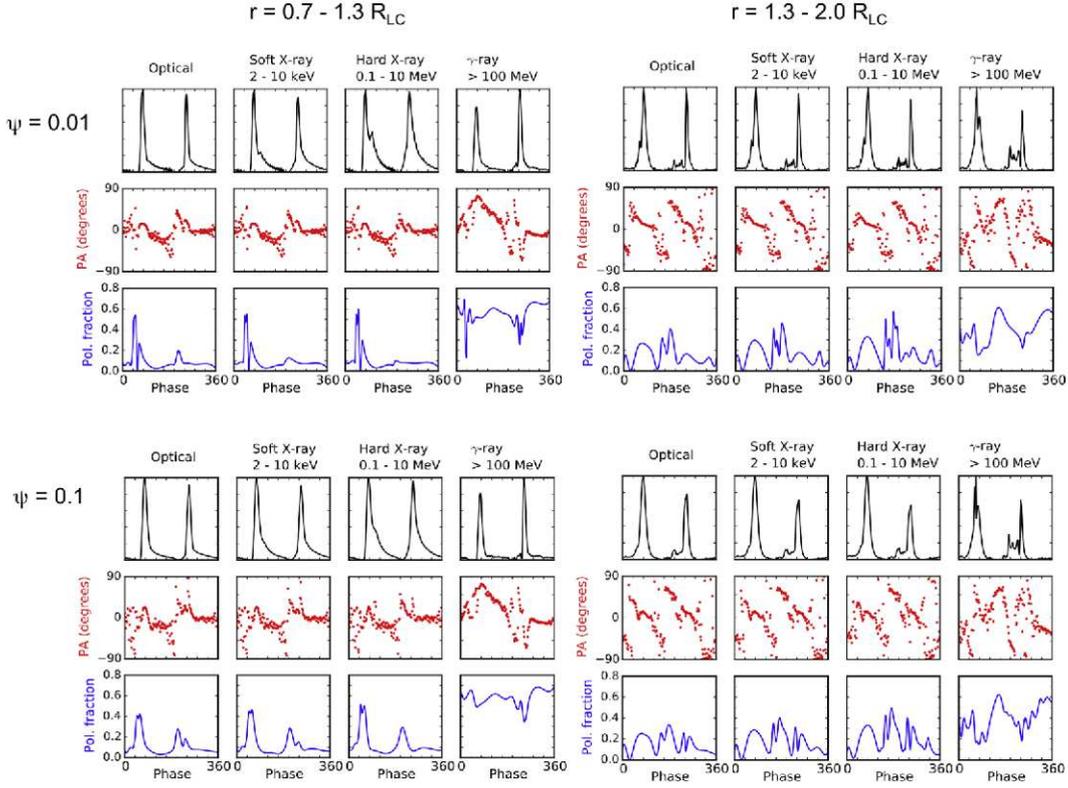}
  \caption{Predictions of light curves (top panels), position angle (middle panels), and polarisation fraction (bottom panels) vs.\ rotation phase for $\alpha= 60^\circ$,
$\zeta = 70^\circ$, and four frequency bands: optical ($1-10$~eV), soft X-ray ($2-10$~keV), hard X-ray ($0.1-10$~MeV), and $\gamma$-ray ($0.1-100$~GeV). Two emission radius ranges were assumed ($r = 0.7-1.3R_{\rm LC}$ and $r = 1.3-2.0R_{\rm LC}$) as well as two constant particle SR pitch angles, $\psi=0.01$ and $\psi=0.1$~\cite{Harding_pol17}.}
  \label{fig:pol}
  \end{center}
\end{figure}
In addition to explaining spectral, light curve, and population features, models should also be able to describe the polarisation signatures that have been or may be seen in pulsars~(e.g., \cite{Kalapotharakos10,Petri05,Takata07}). Thus, polarisation studies provide an additional constraint on $B$-field structure, obliquity and viewing angle, and magnetospheric emission physics while also aiding in model scrutiny and discrimination\footnote{For example, authors have invoked SR~\cite{Petri05}, CR~\cite{Kalapotharakos14} or inverse Compton upscattering~\cite{Aharonian12,Lyutikov13} from beyond the light cylinder to explain the observed $\gamma$-ray signatures. Polarisation properties of these models should be qualitatively different.}.
Dyks \textit{et al.}~\cite{Dyks04} studied the effect of Special Relativity on HE pulsar light curves and polarisation in the TPC, OG, and PC models using a retarded vacuum $B$-field geometry~\cite{Deutsch55}. They found that the TPC could qualitatively reproduce the optical polarisation measurements of the Crab pulsar~\cite{Slowikowska09}. In particular, they found fast swings of the position angle and dips in linear polarisation degree at the phases of the HE light curve peaks. This effect arises from the caustic nature of the HE beam: there is a bunching of photons from different emission altitudes (where local $B$-field vector orientations may differ) in a small observer phase range, leading to a depolarisation as well as rapid position angle changes over this small range in phase. Additionally, there is also a depolarisation due to the superposition of radiation patterns originating from opposite magnetic poles. Cerutti \textit{et al.}~\cite{Cerutti16b} calculated Stokes parameters using their 3D PIC code. They studied HE SR emission originating in the current sheet and found that this emission is mildly polarised (they found an average degree of linear polarisation of $\sim15\%$ in the on-pulse and $\sim30\%$ in the off-pulse intervals), also showing a clear anti-correlation between flux and degree of linear polarisation as a signature of caustic emission (but this time in the current sheet as opposed to interior to the light cylinder), similar to the findings of Dyks \textit{et al.}~\cite{Dyks04}. Cerutti \textit{et al.}~\cite{Cerutti16b} associated the rapid swings in the polarisation angle near the peak phases with the change of $B$-field orientation (polarity) as the observer's line of sight passes through the current sheet. Harding \& Kalapotharakos~\cite{Harding_pol17} calculated multi-frequency polarisation characteristics of pulsar emission invoking emission from the outer FF magnetosphere and current sheet. They assumed that optical to hard X-ray emission is produced by SR from electron-positron pairs and $\gamma$-ray emission is due to either CR or SR from primary electrons. Large swings in position angle coupled with strong depolarisation dips occurred near the light curve peak phases in all energy bands. The SR polarisation characteristics were found to be very sensitive to the photon emission radius: larger position angle swings occurred for emission outside the light cylinder as the line of sight crossed the current sheet. The phase-averaged SR polarisation degree climbed from 10\% to $\sim$20\% for emission inside vs.\ outside the light cylinder. On the other hand, the polarisation degree for CR was up to $40\%-60\%$, with the dips being wider and deeper for emission outside the light cylinder (Figure~\ref{fig:pol}). A sharp increase in polarisation degree together with a change in position angle at the transition between X-ray and $\gamma$-ray spectral components would confirm CR as the $\gamma$-ray emission mechanism.

\section{Conclusion}\label{sec:Conc}
In this paper, we reviewed general observed properties of HE pulsars, indicating the major impact of the \textit{Fermi} LAT, which has confirmed and broadened many important conclusions by its predecessor, \textit{CGRO}. We have also described the standard theoretical framework plus new developments to explain these properties. We summarised new directions that are being pursued, such as including the effect of general relativity on pair production and PC shapes, studying multipolar fields, and making predictions for polarisation signatures expected for different emission mechanisms. Continued development of our technological capabilities, theoretical model development, computational advances, and better data acquisition should aid us in pushing the boundaries of our understanding of the pulsar phenomenon.


\begin{thebibliography}{99}
\bibitem{HESS_PWN17}{H. Abdalla \textit{et al.} 2017, \textit{The Population of TeV Pulsar Wind Nebulae in the H.E.S.S. Galactic Plane Survey}, \textit{arXiv:1702.08280}}
\bibitem{Fermi_Crab}{A.~A. Abdo \textit{et al.} 2010, \textit{Fermi Large Area Telescope Observations of the Crab Pulsar and Nebula}, \textit{ApJ}, \textbf{708}, 1254}
\bibitem{Vela2}{A.~A. Abdo \textit{et al.} 2010, \textit{The Vela Pulsar: Results from the First Year of Fermi LAT Observations}, \textit{ApJ}, \textbf{713}, 154}
\bibitem{Crab_flares}{A.~A. Abdo \textit{et al.} 2011, \textit{Gamma-Ray Flares from the Crab Nebula}, \textit{Science}, \textbf{331}, 739}
\bibitem{B1259}{A.~A. Abdo \textit{et al.} 2011, \textit{Discovery of High-energy Gamma-ray Emission from the Binary System PSR~B1259-63/LS 2883 around Periastron with Fermi}, \textit{ApJL}, \textbf{736}, L11}
\bibitem{2PC}{A.~A. Abdo \textit{et al.} 2013, \textit{The Second Fermi Large Area Telescope Catalog of Gamma-Ray Pulsars}, \textit{ApJS}, \textbf{208}, 17}
\bibitem{2HWC}{A.~U. Abeysekara \textit{et al.} 2017, \textit{The 2HWC HAWC Observatory Gamma-Ray Catalog}, \textit{ApJ}, \textbf{843}, 40}
\bibitem{Acero13a}{F. Acero 2013a, \textit{Two Candidate Radio-quiet Millisecond Pulsars in Fermi Unassociated Sources}, \textit{XMM-Newton Proposal}}
\bibitem{Acero13b}{F. Acero \textit{et al.} 2013b, \textit{Hunting for Treasures among the Fermi Unassociated Sources: A Multiwavelength Approach}, \textit{ApJ}, \textbf{779}, 133}
%\bibitem{1FHL}{M. Ackermann \textit{et al.} 2013, \textit{The First Fermi-LAT Catalog of Sources above 10 GeV}, \textit{ApJS}, \textbf{209}, 34}
%\bibitem{2FHL}{M. Ackermann \textit{et al.} 2016, \textit{2FHL: The Second Catalog of Hard Fermi-LAT Sources}, \textit{ApJS}, \textbf{22}, 5}
\bibitem{Aharonian12}{F.~A. Aharonian, S.~V. Bogovalov, \& D. Khangulyan 2012, \textit{Abrupt Acceleration of a `Cold' Ultrarelativistic Wind from the Crab Pulsar}, \textit{Nature}, \textbf{482}, 507}
\bibitem{3FHL}{M. Ajello \textit{et al.} 2017, \textit{3FHL: The Third Catalog of Hard Fermi-LAT Sources}, \textit{ApJS}, \textbf{232}, 18}
\bibitem{MAGIC_Crab11}{J. Aleksi\'c \textit{et al.} 2011, \textit{Observations of the Crab Pulsar between 25 and 100 GeV with the MAGIC I Telescope}, \textit{ApJ}, \textbf{742}, 43}
\bibitem{MAGIC_Crab12}{J. Aleksi\'c \textit{et al.} 2012, \textit{Phase-resolved Energy Spectra of the Crab Pulsar in the Range of $50-400$ GeV Measured with the MAGIC Telescopes}, \textit{A\&A}, \textbf{540}, A69}
\bibitem{MAGIC_Crab08}{E. Aliu \textit{et al.} 2008, \textit{Observation of Pulsed {$\gamma$}-Rays Above 25 GeV from the Crab Pulsar with MAGIC}, \textit{Science}, \textbf{322}, 1221}
\bibitem{Allafort13}{A. Allafort \textit{et al.} 2013, \textit{PSR J2021+4026 in the Gamma Cygni Region: The First Variable {$\gamma$}-Ray Pulsar Seen by the Fermi LAT}, \textit{ApJL}, \textbf{777}, L2}
\bibitem{Ansoldi16}{S. Ansoldi \textit{et al.} 2016, \textit{Teraelectronvolt Pulsed Emission from the Crab Pulsar detected by MAGIC}, \textit{A\&A}, \textbf{585}, A133}
\bibitem{Archibald16}{R.~F. Archibald \textit{et al.} 2016, \textit{A High Braking Index for a Pulsar}, \textit{ApJL}, \textbf{819}, L16}
\bibitem{Arons79}{J. Arons \& E.~T. Scharlemann 1979, \textit{Pair Formation above Pulsar Polar Caps - Structure of the Low-altitude Acceleration Zone}, \textit{ApJ}, \textbf{231}, 854}
\bibitem{Arons83}{J. Arons 1983, \textit{Pair Creation Above Pulsar Polar Caps - Geometrical Structure and Energetics of Slot Gaps}, \textit{ApJ}, \textbf{266}, 215}
\bibitem{Arzoumanian14}{Z. Arzoumanian \textit{et al.} 2014, \textit{The Neutron Star Interior Composition Explorer (NICER): Mission Definition}, \textit{in: Space Telescopes and Instrumentation 2014: Ultraviolet to Gamma Ray, Proc.\ SPIE 9144}, 914420}
\bibitem{Asseo02}{E. Asseo \& D. Khechinashvili 2002, \textit{The Role of Multipolar Magnetic fields in Pulsar Magnetospheres}, \textit{MNRAS}, \textbf{334}, 743}
\bibitem{Atwood2009}{W.~B. Atwood \textit{et al.} 2009, \textit{The Large Area Telescope on the Fermi Gamma-Ray Space Telescope Mission}, \textit{ApJ}, \textbf{697}, 1071}
\bibitem{BaadeZwicky34}{W. Baade \& F. Zwicky 1934, \textit{Cosmic Rays from Supernovae}, \textit{Proc.\ Nat.\ Acad.\ Sci.}, \textbf{20}, 259}
\bibitem{Bai10}{X.-N. Bai \& A. Spitkovsky 2010, \textit{Uncertainties of Modeling Gamma-ray Pulsar Light Curves Using Vacuum Dipole Magnetic Field}, \textit{ApJ}, \textbf{715}, 1270}
\bibitem{Barnard16}{M. Barnard, C. Venter, \& A.~K. Harding 2016, \textit{The Effect of an Offset Polar Cap Dipolar Magnetic Field on the Modeling of the Vela Pulsar's {$\gamma$}-Ray Light Curves}, \textit{ApJ}, \textbf{832}, 107}
\bibitem{Belyaev15a}{M.~A. Belyaev 2015a, \textit{PICsar: A 2.5D Axisymmetric, Relativistic, Electromagnetic, Particle in Cell Code with a Radiation Absorbing Boundary}, \textit{NewA}, \textbf{36}, 37}
\bibitem{Belyaev15b}{M.~A. Belyaev 2015b, \textit{Dissipation, Energy transfer, and Spin-down Luminosity in 2.5D PIC Simulations of the Pulsar Magnetosphere}, \textit{MNRAS}, \textbf{449}, 2759}
\bibitem{Beskin83}{V.~S. Beskin, A.~V. Gurevich, \& I.~N. Istomin 1983, \textit{The Electrodynamics of a Pulsar Magnetosphere}, \textit{Zhurnal Eksperimentalnoi i Teoreticheskoi Fiziki}, \textbf{85}, 401}
\bibitem{Beskin17}{V.~S. Beskin, A.~K. Galishnikova, E.~M. Novoselov, A.~A. Philippov, \& M.~M. Rashkovetskyi 2017,, \textit{So How Do Radio Pulsars Slow Down?}, \textit{J.\ Phys.\ Conf.\ Ser.}, \textbf{932}, 012012}
\bibitem{Bogdanov2007}{S. Bogdanov, G.~B. Rybicki, \& J.~E. Grindlay 2007, \textit{Constraints on Neutron Star Properties from X-Ray Observations of Millisecond Pulsars}, \textit{ApJ}, \textbf{670}, 668}
\bibitem{Bogdanov2009}{S. Bogdanov \& J.~E. Grindlay 2009, \textit{Deep XMM-Newton Spectroscopic and Timing Observations of the Isolated Radio Millisecond Pulsar PSR J0030+0451}, \textit{ApJ}, \textbf{703}, 1557}
\bibitem{Bonazzola96}{S. Bonazzola \& E. Gourgoulhon 1996, \textit{Gravitational Waves from Pulsars: Emission by the Magnetic-field-induced Distortion.}, \textit{A\&A}, \textbf{312}, 675}
\bibitem{Brambilla15}{G. Brambilla, C. Kalapotharakos, A.~K. Harding, \& D. Kazanas 2015, \textit{Testing Dissipative Magnetosphere Model Light Curves and Spectra with Fermi Pulsars}, \textit{ApJ}, \textbf{804}, 84}
\bibitem{Brambilla18}{G. Brambilla, C. Kalapotharakos, A. Timokhin, A.~K. Harding, \& D. Kazanas 2018, \textit{Electron-positron Pair Flow and Current Composition in the Pulsar Magnetosphere}, \textit{submitted to ApJ (arXiv:1710.03536)}}
\bibitem{Buehler14}{R. B\"uhler \& R. Blandford 2014, \textit{The Surprising Crab Pulsar and its Nebula: a Review}, \textit{Reports on Progress in Physics}, \textbf{77}, 066901}
\bibitem{Burtovoi17}{A. Burtovoi, T.~Y. Saito, L. Zamieri, \& T. Hassan 2017, \textit{Prospects for the Detection of High-energy (E $>25$~GeV) Fermi pulsars with the Cherenkov Telescope Array}, \textit{MNRAS}, \textbf{471}, 431}
\bibitem{Celik11}{O. Celik \& T.~J. Johnson 2011, \textit{On the Phase-Averaged Spectrum of Pulsars and Shape of Their Cutoffs}, \textit{AIP Conf.\ Ser.}, ed. 
M. Burgay, N. D'Amico, P. Esposito, A. Pellizzoni, \& A. Possenti, \textbf{1357}, 225}
\bibitem{Caraveo14}{P. Caraveo 2014, \textit{Gamma-Ray Pulsar Revolution}, \textit{Ann.\ Rev.\ Astron.\ Astrophys.}, \textbf{52}, 211}
\bibitem{Cerutti15}{B. Cerutti, A.~A. Philippov, K. Parfrey, \& A. Spitkovsky 2015, \textit{Particle Acceleration in Axisymmetric Pulsar Current Sheets}, \textit{MNRAS}, \textbf{448}, 606}
\bibitem{Cerutti16a}{B. Cerutti, A.~A. Philippov, \& A. Spitkovsky 2016a, \textit{Modelling High-energy Pulsar Light Curves from First Principles}, \textit{MNRAS}, \textbf{457}, 2401}
\bibitem{Cerutti16b}{B. Cerutti, J. Mortier, \& A.~A. Philippov 2016b, \textit{Polarized Synchrotron Emission from the Equatorial Current Sheet in Gamma-ray Pulsars}, \textit{MNRAS}, \textbf{463}, L89}
\bibitem{Cerutti17a}{B. Cerutti \& A.~M. Beloborodov 2017, \textit{Electrodynamics of Pulsar Magnetospheres}, \textit{Space Sci.\ Rev.}, \textbf{207}, 111}
\bibitem{Cerutti17b}{B. Cerutti \& A.~A. Philippov 2017, \textit{Dissipation of the Striped Pulsar Wind}, \textit{A\&A}, \textbf{607}, A134}
\bibitem{Chen93}{K. Chen \& M. Ruderman 1993, \textit{Pulsar death lines and death valley}, \textit{ApJ}, \textbf{402}, 264}
\bibitem{Chen14}{A.~Y. Chen \& A.~M. Beloborodov 2014, \textit{Electrodynamics of Axisymmetric Pulsar Magnetosphere with Electron-Positron Discharge: A Numerical Experiment}, \textit{ApJL}, \textbf{795}, L22}
\bibitem{CHR86}{K.~S. Cheng, C. Ho, \& M. Ruderman 1986, \textit{Energetic Radiation from Rapidly Spinning Pulsars. I - Outer Magnetosphere Gaps. II - Vela and Crab}, \textit{ApJ}, \textbf{300}, 500}
\bibitem{Clark17}{C.~J. Clark \textit{et al.} 2017, \textit{The Einstein@Home Gamma-ray Pulsar Survey. I. Search Methods, Sensitivity, and Discovery of New Young Gamma-Ray Pulsars}, \textit{ApJ}, \textbf{834}, 106}
\bibitem{CFK99}{I. Contopoulos, D. Kazanas, \& C. Fendt 1999, \textit{The Axisymmetric Pulsar Magnetosphere}, \textit{ApJ}, \textbf{511}, 351}
\bibitem{DH96}{J.~K. Daugherty, \& A.~K. Harding 1996, \textit{Gamma-Ray Pulsars: Emission from Extended Polar Cap Cascades}, \textit{ApJ}, \textbf{458}, 278}
\bibitem{DeAngelis17}{De Angelis, A. 2017, \textit{Science with e-ASTROGAM}, arXiv:1711.01265}
\bibitem{DeNaurois15}{M. de Naurois 2015, \textit{The Very High Energy Sky from  20 GeV to Hundreds of TeV - Selected Highlights}, \textit{Proc. 34th International Cosmic Ray Conference (ICRC2015)}, ed. A.~S. Borisov \textit{et al.}, 21}
\bibitem{Deutsch55}{A.~J. Deutsch 1955, \textit{The Electromagnetic Field of an Idealized Star in Rigid Rotation in Vacuo}, \textit{Ann. d'Astrophys.}, \textbf{18}, 1}
\bibitem{Du11}{Y.~J. Du, J.~L. Han, G.~J. Qiao, C.~K. Chou 2011, \textit{Gamma-ray Emission from the Vela Pulsar Modeled with the Annular Gap and the Core Gap}, \textit{ApJ}, \textbf{731}, 2}
\bibitem{DR03}{J. Dyks \& B. Rudak 2003, \textit{Two-Pole Caustic Model for High-Energy Light Curves of Pulsars}, \textit{ApJ}, \textbf{598}, 1201}
\bibitem{Dyks04}{J. Dyks, A.~K. Harding, \& B. Rudak 2004, \textit{Relativistic Effects and Polarization in Three High-Energy Pulsar Models}, \textit{ApJ}, \textbf{606}, 1125}
\bibitem{Gil2001}{J. Gil \& D. Mitra 2001, \textit{Vacuum Gaps in Pulsars and PSR J2144-3933}, \textit{ApJ}, \textbf{550}, 383}
\bibitem{GJ69}{P. Goldreich \& W.~H. Julian 1969, \textit{Pulsar Electrodynamics}, \textit{ApJ}, \textbf{157}, 869}
\bibitem{Gonthier04}{P.~L. Gonthier, R. van Guilder, \& A.~K. Harding 2004, \textit{Role of Beam Geometry in Population Statistics and Pulse Profiles of Radio and Gamma-Ray Pulsars}, \textit{ApJ}, \textbf{604}, 775}
\bibitem{Gonthier07}{P.~L. Gonthier, S.~A. Story, B.~D. Clow, \& A.~K. Harding 2007, \textit{Population Statistics Study of Radio and Gamma-ray Pulsars in the Galactic Plane}, \textit{A\&SS}, \textbf{309}, 245}
\bibitem{Gralla16}{S.~E. Gralla, A. Lupsasca, \& A. Philippov 2016, \textit{Pulsar Magnetospheres: Beyond the Flat Spacetime Dipole}, \textit{ApJ}, \textbf{833}, 258}
\bibitem{Gralla17}{S.~E. Gralla, A. Lupsasca, \& A. Philippov 2017, \textit{Inclined Pulsar Magnetospheres in General Relativity: Polar Caps for the Dipole, Quadrudipole and Beyond}, \textit{ApJ}, \textbf{851}, 137}
\bibitem{Grenier15}{I. Grenier \& A.~K. Harding 2015, \textit{Gamma-ray Pulsars: A Gold Mine}, \textit{Comptes Rendus Physique}, \textbf{16}, 641}
\bibitem{Guillemot14}{L. Guillemot \& T.~M. Tauris 2014, \textit{On the Non-detection of {$\gamma$}-rays from Energetic Millisecond Pulsars - Dependence on Viewing Geometry}, \textit{MNRAS}, \textbf{439}, 2033}
\bibitem{Guillemot16}{L. Guillemot \textit{et al.} 2016, \textit{The Gamma-ray Millisecond Pulsar Deathline, Revisited. New Velocity and Distance Measurements}, \textit{A\&A}, \textbf{587}, A109}
\bibitem{Han97}{J.~L. Han 1997, \textit{Slowly Rotating Pulsars and Magnetic Field Decay}, \textit{A\&A}, \textbf{318}, 485}
\bibitem{Harding81}{A.~K. Harding 1981, \textit{Pulsar Gamma rays - Spectra, Luminosities, and Efficiencies}, \textit{ApJ}, \textbf{245}, 267}
\bibitem{HUM05}{A.~K. Harding, V.~V. Usov, \& A.~G. Muslimov 2005, \textit{High-Energy Emission from Millisecond Pulsars}, \textit{ApJ}, \textbf{622}, 531}
\bibitem{HM11a}{A.~K. Harding \& A.~G. Muslimov 2011a, \textit{Pulsar Pair Cascades in a Distorted Magnetic Dipole Field}, \textit{ApJL}, \textbf{726}, L10}
\bibitem{HM11b}{A.~K. Harding \& A.~G. Muslimov 2011b, \textit{Pulsar Pair Cascades in Magnetic Fields with Offset Polar Caps}, \textit{ApJ}, \textbf{743}, 181}
\bibitem{Harding15}{A.~K. Harding \& C. Kalapotharakos 2015, \textit{Synchrotron Self-Compton Emission from the Crab and Other Pulsars}, \textit{ApJ}, \textbf{811}, 63}
\bibitem{Harding17}{A.~K. Harding \& C. Kalapotharakos 2017a, \textit{MeV Pulsars: Modeling Spectra and Polarization}, in PoS, Proc.\ 7$^{\rm th}$ International Fermi Symp.\ (IFS2017)006, arXiv:1712.02406}
\bibitem{Harding_pol17}{A.~K. Harding \& C. Kalapotharakos 2017b, \textit{Multiwavelength Polarization of Rotation-powered Pulsars}, \textit{ApJ}, \textbf{840}, 73}
\bibitem{Igoshev15}{A.~P. Igoshev \& S.~B. Popov 2015, \textit{Magnetic Field Decay in Normal Radio Pulsars}, \textit{AN}, \textbf{336}, 831}
\bibitem{Johnson14}{T.~J. Johnson \textit{et al.} 2014, \textit{Constraints on the Emission Geometries and Spin Evolution of Gamma-Ray Millisecond Pulsars}, \textit{ApJS}, \textbf{213}, 6}
\bibitem{Aris17}{S. Johnston \& A. Karastergiou 2017, \textit{Pulsar Braking and the P-$\dot{P}$ Diagram}, \textit{MNRAS}, \textbf{467}, 3493}
\bibitem{Kalapotharakos10}{C. Kalapotharakos \& I. Contopoulos 2010, \textit{The Pulsar Synchrotron in 3D: Curvature Radiation}, \textit{MNRAS}, \textbf{404}, 767}
\bibitem{Kalapotharakos12}{C. Kalapotharakos, D. Kazanas, A.~K. Harding, \& I. Contopoulos 2012, \textit{Toward a Realistic Pulsar Magnetosphere}, \textit{ApJ}, \textbf{749}, 2}
\bibitem{Kalapotharakos14}{C. Kalapotharakos, A.~K. Harding, \& D. Kazanas 2014, \textit{Gamma-Ray Emission in Dissipative Pulsar Magnetospheres: From Theory to Fermi Observations}, \textit{ApJ}, \textbf{793}, 97}
\bibitem{Kalapotharakos17a}{C. Kalapotharakos, A.~K. Harding, D. Kazanas, \& G. Brambilla 2017, \textit{Fermi Gamma-Ray Pulsars: Understanding the High-energy Emission from Dissipative Magnetospheres}, \textit{ApJ}, \textbf{842}, 80}
\bibitem{Kalapotharakos17b}{C. Kalapotharakos, G. Brambilla, A. Timokhin, A.~K. Harding, \& D. Kazanas 2018, \textit{3D Kinetic Pulsar Magnetosphere Models: Exploring Self Consistency}, \textit{submitted to ApJ (arXiv:1710.03170)}}
\bibitem{Komissarov07}{S.~S. Komissarov 2007, \textit{Multidimensional Numerical Scheme for Resistive Relativistic Magnetohydrodynamics}, \textit{MNRAS}, \textbf{382}, 995}
\bibitem{Kong14}{A. Kong 2014, \textit{X-ray Observations of Radio-quiet Black Widow-type Millisecond Pulsars}, \textit{in: The X-ray Universe 2014}, ed. J.-U. Ness, 106}
\bibitem{Kuiper15}{L. Kuiper \& W. Hermsen 2015, \textit{The Soft {$\gamma$}-ray Pulsar Population: a High-energy Overview}, \textit{MNRAS}, \textbf{449}, 3827}
\bibitem{Kundu17}{A. Kundu \& J. P\'etri 2017, \textit{Pulsed Emission from a Rotating Off-centred Magnetic Dipole in Vacuum}, \textit{MNRAS}, \textbf{471}, 3359}
\bibitem{Kuzmin86}{A.~D. Kuzmin, V.~M. Malofeev, V.~A. Izvekova, W. Sieber, \& R. Wielebinski 1986, \textit{A comparison of high-frequency and low-frequency characteristics of pulsars}, \textit{A\&A} \textbf{161}, 183}
\bibitem{Leung14}{G.~C.~K. Leung, J. Takata, C.~W. Ng, A.~K.~H. Kong, P.~H.~T. Tam, C.~Y. Hui, \& K.~S. Cheng 2014, \textit{Fermi-LAT Detection of Pulsed Gamma-Rays above 50 GeV from the Vela Pulsar}, \textit{ApJL}, \textbf{797}, L13}
\bibitem{Li12}{J. Li, A. Spitkovsky, \& A. Tchekhovskoy 2012, \textit{Resistive Solutions for Pulsar Magnetospheres}, \textit{ApJ}, \textbf{746}, 60}
\bibitem{Lyne15}{A.~G. Lyne, B.~W. Stappers, M.~J. Keith, P.~S. Ray, M. Kerr, F. Camilo, \& T.~J. Johnson 2015, \textit{The Binary Nature of PSR~J2032+4127}, \textit{MNRAS}, \textbf{451}, 581}
\bibitem{Lyutikov13}{M. Lyutikov 2013, \textit{Inverse Compton Model of Pulsar High-energy Emission}, \textit{MNRAS}, \textbf{431}, 2580}
\bibitem{McCann15}{A. McCann 2015, \textit{A Stacked Analysis of 115 Pulsars Observed by the FERMI LAT}, \textit{ApJ}, \textbf{804}, 86}
\bibitem{Michel91}{F.~C. Michel 1991, \textit{Theory of Neutron Star Magnetospheres}, University of Chicago Press}
\bibitem{Moiseev2017}{A. Moiseev \textit{et al.}, \emph{All-Sky Medium Energy Gamma-ray Observatory (AMEGO)}, \textit{ICRC2017}, PoS, \textbf{301}, 798}
\bibitem{Montgomery99}{H. Montgomery 1999, \textit{Unipolar Induction: a Neglected Topic in the Teaching of Electromagnetism}, \textit{Eur.\ J.\ Phys.}, \textbf{20}, 271}
\bibitem{Ostriker69}{J.~P. Ostriker \& J.~E. Gunn 1969, \textit{On the Nature of Pulsars. I. Theory}, \textit{ApJ}, \textbf{157}, 1395}
\bibitem{Pandharipande76}{V.~R. Pandharipande, D. Pines, \& R.~A. Smith 1976, \textit{Neutron Star Structure: Theory, Observation, and Speculation}, \textit{ApJ}, \textbf{208}, 550}
\bibitem{Petri05}{J. P{\'e}tri, \& J.~G. Kirk 2005, \textit{The Polarization of High-Energy Pulsar Radiation in the Striped Wind Model}, \textit{ApJL}, \textbf{627}, L37}
\bibitem{Petri11}{J. P\'etri 2011, \textit{A Unified Polar Cap/Striped Wind Model for Pulsed Radio and Gamma-ray Emission in Pulsars}, \textit{MNRAS}, \textbf{412}, 1870}
\bibitem{Petri12}{J. P\'etri 2012, \textit{High-energy Emission from the Pulsar Striped Wind: a Synchrotron Model for Gamma-ray Pulsars}, \textit{MNRAS}, \textbf{424}, 2023}
\bibitem{Petri15}{J. P\'etri 2015, \textit{Multipolar Electromagnetic Fields around Neutron Stars: Exact Vacuum Solutions and Related Properties}, \textit{MNRAS}, \textbf{450}, 714}
\bibitem{Petri16}{J. P\'etri 2016, \textit{Radiation from an Off-centred Rotating Dipole in Vacuum}, \textit{MNRAS}, \textbf{463}, 1240}
\bibitem{Petri17a}{J. P\'etri 2017a, \textit{Polarized emission from an off-centred dipole}, \textit{MNRAS}, \textbf{466}, L73}
\bibitem{Petri17b}{J. P\'etri 2017b, \textit{Multipolar Electromagnetic Fields around Neutron Stars: General-relativistic Vacuum Solutions}, \textit{MNRAS}, \textbf{472}, 3304}
\bibitem{Philippov14}{A.~A. Philippov \& A. Spitkovsky 2014, \textit{Ab Initio Pulsar Magnetosphere: Three-dimensional Particle-in-cell Simulations of Axisymmetric Pulsars}, \textit{ApJL}, \textbf{785}, L33}
\bibitem{Philippov15a}{A.~A. Philippov, A. Spitkovsky, \& B. Cerutti 2015a, \textit{Ab Initio Pulsar Magnetosphere: Three-dimensional Particle-in-cell Simulations of Oblique Pulsars}, \textit{ApJL}, \textbf{801}, L19}
\bibitem{Philippov15b}{A.~A. Philippov, B. Cerutti, A. Tchekhovskoy, \& A. Spitkovsky 2015b, \textit{Ab Initio Pulsar Magnetosphere: The Role of General Relativity}, \textit{ApJL}, \textbf{815}, L19}
\bibitem{Philippov17}{A.~A. Philippov \& A. Spitkovsky 2018, \textit{Ab-Initio Pulsar Magnetosphere: Particle Acceleration in Oblique Rotators and High-energy Emission Modeling}, \textit{submitted to ApJ (arXiv:1707.04323)}}
\bibitem{Pierbattista15}{M. Pierbattista, A.~K. Harding, I.~A. Grenier, T.~J. Johnson, P.~A. Caraveo, M. Kerr, \& P.~L. Gonthier 2015, \textit{Light-curve Modelling Constraints on the Obliquities and Aspect Angles of the Young Fermi Pulsars}, \textit{A\&A}, \textbf{575}, A3}
\bibitem{Pletsch12}{H.~J. Pletsch \textit{et al.} 2012, \textit{Discovery of Nine Gamma-Ray Pulsars in Fermi Large Area Telescope Data Using a New Blind Search Method}, \textit{ApJ}, \textbf{744}, 105}
\bibitem{Qiao04}{G.~J. Qiao, K.~J. Lee, H.~G. Wang, R.~X. Xu J.~L. Han 2004, \textit{The Inner Annular Gap for Pulsar Radiation: {$\gamma$}-Ray and Radio Emission}, \textit{ApJ}, \textbf{606}, L49}
\bibitem{Renault15}{N. Renault-Tinacci, I. Grenier, \& A.~K. Harding 2015, \textit{Phase-Resolved Spectral Analysis of 25 Millisecond Gamma-ray Pulsars}, \textit{34th International Cosmic Ray Conference (ICRC2015)}, \textbf{34}, 843}
\bibitem{Roberts11}{M.~S.~E. Roberts 2011, \textit{New Black Widows and Redbacks in the Galactic Field}, \textit{AIP Conf.\ Ser.}, ed. 
M. Burgay, N. D'Amico, P. Esposito, A. Pellizzoni, \& A. Possenti, \textbf{1357}, 127}
\bibitem{RY95}{R. Romani \& I.-A. Yadigaroglu 1995, \textit{Gamma-ray Pulsars: Emission Zones and Viewing Geometries}, \textit{ApJ}, \textbf{438}, 314}
\bibitem{R96}{R. Romani 1996, \textit{Gamma-Ray Pulsars: Radiation Processes in the Outer Magnetosphere}, \textit{ApJ}, \textbf{470}, 469}
\bibitem{RS75}{M.~A. Ruderman \& P.~G. Sutherland 1975, \textit{Theory of Pulsars - Polar Caps, Sparks, and Coherent Microwave Radiation}, \textit{ApJ}, \textbf{196}, 51}
\bibitem{SazParkinson2010}{P.~M. Saz Parkinson \textit{et al.} 2010, \textit{Eight {$\gamma$}-ray Pulsars Discovered in Blind Frequency Searches of Fermi LAT Data}, \textit{ApJ}, \textbf{725}, 571}
\bibitem{Schmidt05}{P. Schmidt \textit{et al.} 2005, \textit{Search for Pulsed TeV Gamma-Ray Emission from Young Pulsars with H.E.S.S.}, \textit{AIP Conf.\ Ser.}, ed. F.~A. Aharonian, H.~J. V\"olk, \& D. Horns, 377}
\bibitem{Slowikowska09}{A. S{\l}owikowska, G. Kanbach, M. Kramer, \& A. Stefanescu 2009, \textit{Optical Polarization of the Crab Pulsar: Precision Measurements and Comparison to the Radio Emission}, \textit{MNRAS}, \textbf{397}, 103}
\bibitem{Smith17}{D.~A. Smith, L. Guillemot, M. Kerr, C. Ng, \& E. Barr  2017, \textit{Gamma-ray Pulsars with \textit{Fermi}}, \textit{arXiv:1706.03592}}
\bibitem{Spitkovsky2006}{A. Spitkovsky 2006, \textit{Time-dependent Force-free Pulsar Magnetospheres: Axisymmetric and Oblique Rotators}, \textit{ApJL}, \textbf{648}, L51}
\bibitem{Story14}{S.~A. Story \& M~G. Baring 2014, \textit{Magnetic Pair Creation Transparency in Gamma-Ray Pulsars}, \textit{ApJ}, \textbf{790}, 61}
\bibitem{Takata07}{J. Takata, \& H.-K. Chang 2007, \textit{Pulse Profiles, Spectra, and Polarization Characteristics of Nonthermal Emissions from the Crab-like Pulsars}, \textit{ApJ}, \textbf{670}, 677}
\bibitem{Takata08}{J. Takata, H.~K. Chang, \& S. Shibata 2008, \textit{Particle Acceleration and Non-thermal Emission in the Pulsar Outer Magnetospheric Gap}, \textit{MNRAS}, \textbf{386}, 748}
\bibitem{Tauris01}{T.~M. Tauris \& S. Konar 2001, \textit{Torque Decay in the Pulsar (P,$\dot{P}$) Diagram. Effects of Crustal Ohmic Dissipation and Alignment}, \textit{A\&A}, \textbf{376}, 543}
\bibitem{AGILE_flares}{M. Tavani \textit{et al.} 2011, \textit{Discovery of Powerful Gamma-Ray Flares from the Crab Nebula}, \textit{Science}, \textbf{331}, 736}
\bibitem{Tchekhovskoy13}{A. Tchekhovskoy, A. Spitkovsky, \& J.~G. Li 2013, \textit{Time-dependent 3D Magnetohydrodynamic Pulsar Magnetospheres: Oblique Rotators}, \textit{MNRAS}, \textbf{435}, L1}
\bibitem{Thompson2004}{D.~J. Thompson 2004, \textit{Gamma-ray Pulsars}, \textit{in: Cosmic Gamma-ray Sources, Astrophysics and Space Science Library}, ed. K.~S. Cheng \& G.~E. Romero,  \textbf{304}, 149}
\bibitem{Timokhin06}{A.~N. Timokhin 2006, \textit{On the Force-Free Magnetosphere of an Aligned Rotator}, \textit{MNRAS}, \textbf{368}, 1055}
\bibitem{Valone01}{T. Valone 2001, \textit{The Homopolar Handbook: A Definitive Guide to Faraday Disk and N-Machine Technologies}, Washington, DC: Integrity Research Institute}
\bibitem{Venter07}{C. Venter 2007, \textit{Constraints on the Parameters of the Unseen Pulsar in the PWN G0.9+0.1 from Radio, X-Ray, and VHE Gamma-Ray Observations}, \textit{in: WE-Heraeus Seminar on Neutron Stars and Pulsars 40 years after the Discovery}, ed. W. Becker \& H.~H. Huang,  40}
\bibitem{Venter09}{C. Venter, A.~K. Harding, \& L. Guillemot 2009, \textit{Probing Millisecond Pulsar Emission Geometry Using Light Curves from the Fermi Large Area Telescope}, \textit{ApJ}, \textbf{707}, 800}
\bibitem{Venter12}{C. Venter, T.~J. Johnson, \& A.~K. Harding 2012, \textit{Modeling Phase-aligned Gamma-Ray and Radio Millisecond Pulsar Light Curves}, \textit{ApJ}, \textbf{744}, 34}
\bibitem{Venter14}{C. Venter \& A.~K. Harding 2014, \textit{High-energy Pulsar Models: Developments and New Questions}, \textit{Astronomische Nachrichten}, \textbf{335}, 268}
\bibitem{Venter15}{C. Venter, A. Kopp, A.~K. Harding, P.~L. Gonthier, \& I. B\"usching 2015, \textit{Cosmic-ray Positrons from Millisecond Pulsars}, \textit{ApJ}, \textbf{807}, 130}
\bibitem{Venter16_HEASA}{C. Venter 2016, \textit{New Advances in the Modelling of Pulsar Magnetospheres}, \textit{Proc.\ 4th Annual Conference on High Energy Astrophysics in Southern Africa (HEASA 2016)}, 40 (http://pos.sissa.it/cgi-bin/reader/conf.cgi?confid=275, id.40)}
\bibitem{Watters09}{K.~P. Watters, R.~W. Romani, P. Weltevrede, \& S. Johnston 2009, \textit{An Atlas for Interpreting $\gamma$-Ray Pulsar Light Curves}, \textit{ApJ}, \textbf{695}, 1298}
\bibitem{Watters11}{K.~P. Watters \& R.~W. Romani 2011, \textit{The Galactic Population of Young {$\gamma$}-ray Pulsars}, \textit{ApJ}, \textbf{727}, 123}
\bibitem{Xing16}{Y. Xing, Z. Wang, \& J. Takata 2016, \textit{Fermi Study of 5--300 GeV Emission from the High-mass Pulsar Binary PSR B1259-63/LS 2883}, \textit{ApJ}, \textbf{828}, 61}
\bibitem{Zimmerman80}{M. Zimmerman 1980, \textit{Gravitational Waves from Rotating and Precessing Rigid Bodies. II - General Solutions and Computationally Useful Formulas}, \textit{Phys.\ Rev.\ D}, \textbf{21}, 891}
\end{thebibliography}
\end{document}